\documentclass[manuscript,screen, authoryear]{acmart}
\AtBeginDocument{%
  }

\setcopyright{acmlicensed}
\copyrightyear{2025}
\acmYear{2025}
\acmDOI{XXXXXXX.XXXXXXX}
\acmConference[Conference acronym 'XX]{Make sure to enter the correct
  conference title from your rights confirmation email}{June 03--05,
  2018}{Woodstock, NY}
\acmISBN{978-1-4503-XXXX-X/2018/06}

\usepackage{graphicx}
\usepackage{algorithm}
\usepackage{algpseudocode}
\usepackage{physics}
\usepackage{geometry}

\usepackage{float}
\usepackage{subcaption}
\usepackage{hyperref}

\usepackage{listings}
\usepackage{xcolor}
\usepackage{environ}

\definecolor{codegreen}{rgb}{0,0.6,0}
\definecolor{codegray}{rgb}{0.5,0.5,0.5}
\definecolor{codepurple}{rgb}{0.58,0,0.82}
\definecolor{backcolour}{rgb}{0.95,0.95,0.92}

\lstdefinestyle{mystyle}{
    commentstyle=\color{codegreen},
    keywordstyle=\color{magenta},
    numberstyle=\tiny\color{codegray},
    stringstyle=\color{codepurple},
    basicstyle=\ttfamily\footnotesize,
}
\begin{document}

\title{OptimizedDP: An Efficient, User-friendly Library For Optimal Control and Dynamic Programming}

\author{Minh Bui}
\email{minh_bui_3@sfu.ca}
\affiliation{%
  \institution{Simon Fraser University}
  \city{Burnaby}
  \state{BC}
  \country{Canada}
}

\author{Hanyang Hu}
\email{hha160@sfu.ca}
\affiliation{%
  \institution{Simon Fraser University}
  \city{Burnaby}
  \state{BC}
  \country{Canada}
}

\author{Chong He}
\email{chong_he@sfu.ca}
\affiliation{%
  \institution{Simon Fraser University}
  \city{Burnaby}
  \state{BC}
  \country{Canada}
}

\author{Michael Lu}
\email{mla233@sfu.ca}
\affiliation{%
  \institution{Simon Fraser University}
  \city{Burnaby}
  \state{BC}
  \country{Canada}
}
\author{George Giovanis}
\email{georgedgiovani.dev@gmail.com}
\affiliation{%
 \institution{Amazon}
 \city{Vancouver}
 \state{BC}
 \country{Canada}
}

\author{Arrvindh Shriraman}
\email{arrvindh_shriraman@sfu.ca}
\affiliation{%
  \institution{Simon Fraser University}
  \city{Burnaby}
  \state{BC}
  \country{Canada}
}

\author{Mo Chen}
\email{mochen@sfu.ca}
\affiliation{%
  \institution{Simon Fraser University}
  \city{Burnaby}
  \state{BC}
  \country{Canada}
}

\renewcommand{\shortauthors}{Minh et al.}

\begin{abstract}
This paper introduces OptimizedDP, a high-performance software library for several
common grid-based dynamic programming (DP) algorithms used in control theory and
robotics.
Specifically, OptimizedDP provides functions to numerically solve a class of
time-dependent (dynamic) Hamilton-Jacobi (HJ) partial differential equations (PDEs),
time-independent (static) HJ PDEs, and additionally value iteration for continuous
action-state space Markov Decision Processes (MDP).
The computational complexity of grid-based DP is exponential with respect to the number
of grid or state space dimensions, and thus can have bad execution runtimes and memory usage when
applied to large state spaces.
We leverage the user-friendliness of Python for different problem specifications
without sacrificing the efficiency of the core computation.
This is achieved by implementing the core part of the code which the user does not see
in heterocl, a framework we use to abstract away details of how computation is
parallelized.
Compared to similar toolboxes for level set methods that are used to solve the HJ PDE,
our toolbox makes solving the PDE at higher dimensions possible as well as achieving 
an order of magnitude improvements in execution times, while
keeping the interface easy for specifying different problem descriptions.
Because of that, the toolbox has been adopted to solve control and optimization
problems that were considered intractable before.
Our toolbox is available publicly at \url{https://github.com/SFU-MARS/optimized\_dp}.
\end{abstract}




\maketitle

\section{Introduction}
\label{intro}

Dynamic programming (DP) is central to many control and optimization applications. Despite its 
exponential complexity, globally optimal solutions to many control and optimization problems are only possible via
DP. It also serves as a baseline to which approximative and analytical solutions
can be compared against \cite{Bertsekas2000DP}. In continuous domains, DP is applied by discretizing state, action, 
and time spaces; finer discretization increases accuracy but leads to exponential complexity 
in computation and memory.
While this is a powerful and general approach,
the exponential complexity renders slow running time for modest dimensional problem (4-5 dimensional) 
and intractable for high-dimensional (above 5-dimensional) problems \cite{HJOverview}.

In this paper, we propose OptimizedDP, a software toolbox designed to alleviate the long execution times of DP 
and enable tractable computation for high-dimensional problems by efficiently implementing common grid-based 
algorithms: continuous Markov Decision Process (MDP) value iteration and level-set based methods for
solving time-dependent and time-independent Hamilton Jacobi (HJ) Partial Differential Equations (PDEs). Unlike existing libraries to solve 
MDPs such as POMDP \cite{JuliaMDP} and MDP Toolbox for Python \cite{MDPToolbox}, 
OptimizedDP supports value iteration on continuous state and action spaces through discretization. The level-set methods
for solving HJ PDEs crucially provide solutions to optimal control problems with applications in differential games \cite{HJDynamicGame,HuangRA,Fisac2014ReachavoidPW}
, trajectory planning \cite{Chen2021FaSTrackAMF}, aerial refueling \cite{HJRefuel}, and safety verification 
via reachability analysis \cite{HJOverview,HJOverview2}, with potential broader impact in computer graphics,
fluid dynamics, and beyond \cite{LevesetMacrhing,Osher2002LevelSM}.

Solving the HJ PDE is computationally demanding, requiring complex numerical algorithms and extensive floating-point 
operations on high-dimensional grids. This makes implementing algorithms, prototyping dynamics, and validating
results slow and cumbersome. Several toolboxes have been developed to ease this process: HelperOC 
(a wrapper of ToolboxLS \cite{LsetToolbox1}), hj\_reachability \cite{hjreachability}, and BEACLS \cite{BEACLS}.
ToolboxLS and HelperOC, which are written in MATLAB, provide powerful visualization tools and easy prototyping but suffer
from slow runtimes, proprietary licenses, and low-dimensional scalability. hj\_reachability, which was written 
in Python with JAX, achieves faster execution and provides GPU support but remains limited to small
dimensional problems.
BEACLS, implemented in C++ with GPU support, runs much faster but has a difficult interface for problem specification, making prototyping a bottleneck. Despite fast execution for
small and medium-sized problems, both BEACLS and hj\_reachability face GPU memory limits, preventing their
use from solving high-dimensional problems.

Our toolbox OptimizedDP addresses the shortcomings of the existing toolboxes by
significantly improving the execution runtime and scaling computations to higher dimensions on multi-core CPUs
while keeping the user-friendly interface for problem
specifications. In particular, our contributions are as follows:
\begin{itemize}
    \item Implementation of dynamic programming based algorithms to solve time-dependent Hamilton
    -Jacobi PDEs based on level-set methods \cite{Osher2002LevelSM},
time-independent HJ PDE based on Lax-Friedrich sweeping \cite{TTR}, and value iterations for continuous MDP with continuous state and action space
    \item Efficient implementation of these algorithms that speeds up computational time of up to an order of magnitude compared to
existing toolboxes and allows grid-based DP to be done on grids of up to eight dimensions.
    \item Fast and easy problem instance specification primarily in Python, while the backend implementing and optimizing the algorithms solver is written in HeteroCL, a python-based domain-specific language (DSL) \cite{heterocl}.
\end{itemize}

While we recognize that many recent methods, including deep learning-based \cite{DeepReach} and Hopf-Lax-based
methods \cite{Lax-Hopf} can be more scalable, we believe our toolbox is still valuable because they can serve as a ground truth value functions
for general controlled non-linear systems experiencing disturbances. It can also be a staging ground for 
many methods that scale better when starting from a partial solution or approximate solutions in subspaces.
The toolbox has been used extensively in research community \cite{citedOdp1,citedOdp2,citedOdp3} and enable analysis
of previously intractable control problem \cite{2v1RAGame,UnderwaterPaper}.
Our toolbox is available online at \textbf{\url{https://github.com/SFU-MARS/optimized\_dp}}.

\section{ Overview of Algorithms Supported}

\begin{figure}[H]
    \centering
    \includegraphics[width=0.35\linewidth]{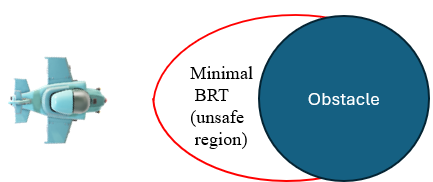}
    \caption{Obtaining Minimal Backward Reachable Tube is crucial for guaranteeing safety.
        The Tube contains all the states the system will inevitable arrive at target set
        despite applying optimal control to avoid.
    }
    \label{fig:BRT_illustration}
\end{figure}
\subsection{Time-dependent (dynamic) Hamilton-Jacobi (HJ) Partial Differential Equation (PDE) }
Solving Hamilton-Jacobi (HJ) Partial Differential Equation (PDE) is a crucial pillar in
reachability analysis and differential games
\cite{HJDynamicGame,Evans1983DifferentialGA,HJOverview} .
In this section, we will first introduce the definitions of Backward Reachable Set
(BRS) and Backward Reachable Tube (BRT), which are important concepts in reachability
analysis and differential games.
Then we will explain how they can be obtained by solving the HJ PDE via level-set based numerical algorithms,
whose optimized implementations are provided in our toolbox.

Let $s \le 0$ be the time and $z \in \mathbb{R}^n$ be the state of a dynamical system, whose evolution is described by a system of ordinary differential equations (ODE) as follows:
\begin{equation}
    \label{eq:main_ode}
    \dot{z} = \dv{z(s)}{s} = f(z(s), u(s), d(s)), u(s) \in \mathcal{U}, d(s) \in \mathcal{D}, s \in [t, 0]
\end{equation}
where $u(\cdot)$ and $d(\cdot)$ respectively denote the control and disturbance function drawn from the set of measurable functions:
\begin{equation}
    u(\cdot) \in \mathbb{U} := \{ \phi : [t, 0] \rightarrow \mathcal{U}, \phi(\cdot) \text{ is measurable}\}
\end{equation}
\begin{equation}
    d(\cdot) \in \mathbb{D} := \{ \phi : [t, 0] \rightarrow \mathcal{D}, \phi(\cdot) \text{ is measurable}\}
\end{equation}
with $\mathcal{U} \subseteq \mathbb{R}^{n_u}$,  $\mathcal{D} \subseteq \mathbb{R}^{n_d}$  are compact and $t < 0$.
The system dynamics $f: \mathbb{R}^n \cross \mathcal{U} \cross \mathcal{D} \rightarrow
    \mathbb{R}^n$ are assumed to be uniformly continuous, bounded and Lipschitz continuous
in $z$ for fixed $u(\cdot)$ and $d(\cdot)$.
The trajectory of the system as the function of time $s$ is denoted as $\zeta(s;z, t,
    u(\cdot), d(\cdot)): [t, 0] \rightarrow \mathbb{R}^n$, which starts from state $z$ at
time $t$ and is under the effects of control function $u(\cdot)$ and disturbances
function $d(\cdot)$.
$\zeta$ satisfies ODE \eqref{eq:main_ode} almost everywhere with initial condition $\zeta(t;z, t, u(\cdot), d(\cdot)) = z$.
In addition, for every $u(\cdot)$ and $d(\cdot)$, there exists a unique trajectory
$\zeta$ that solves equation \eqref{eq:main_ode} \cite{nla.cat-vn2169453}.

From here, we can define \textbf{Minimal Backward Reachable Set (BRS)} as follow:
\begin{equation}
    \label{eq:brs}
    \begin{aligned}
        \mathcal{A} = \{z:  \exists d(\cdot) \in \mathbb{D}, \forall u(\cdot) \in \mathbb{U}, \zeta(0;z, t, u(\cdot), d(\cdot)) \in \mathcal{T} \}
    \end{aligned}
\end{equation}
where $\mathcal{T} \subseteq \mathbb{R}^n $ is the target set, described by an implicit function $\mathcal{T} = \{z: l(z) \leq 0\} $.
Semantically $\mathcal{A}$ can be interpreted as the set of states where the control
system is guaranteed to arrive undesired $\mathcal{T}$ at time $t$ seconds despite the
best control.
For safety-critical application, we would like our system to always avoid arriving
unsafe states such as hitting obstacles at all time.
In such cases,  \textbf{Minimal Backward Reachable Tube (BRT)} can be more useful:
\begin{equation}
    \label{eq:brt}
    \begin{aligned}
        \bar{\mathcal{A}} = \{z:  \exists d(\cdot) \in \mathbb{D}, \forall u(\cdot) \in \mathbb{U}, \exists s \in [t, 0], \zeta(s;z, t, u(\cdot), d(\cdot)) \in \mathcal{T} \}
    \end{aligned}
\end{equation}

This is the set of states from which the system is guaranteed to arrive at
$\mathcal{T}$ \textbf{within} $t$ seconds despite the best control.
Similarly, for reaching, we can define the set of states where the systems would like to arrive, namely \textbf{Maximal Backward Reachable Set} and \textbf{Tubes}:

\begin{equation}
    \mathcal{R} = \{z:  \forall d(\cdot) \in \mathbb{D}, \exists u(\cdot) \in \mathbb{U}, \zeta(0;z, t, u(\cdot), d(\cdot)) \in \mathcal{T} \}
\end{equation}
\begin{equation}
    \bar{\mathcal{R}} = \{z:  \forall d(\cdot) \in \mathbb{D}, \exists u(\cdot) \in \mathbb{U}, \exists s \in [t, 0], \zeta(s;z, t, u(\cdot), d(\cdot)) \in \mathcal{T} \}
\end{equation}
In the case of \textbf{BRS} consider the following function:

\begin{equation}
    \phi(z, t) = \max_{u(\cdot)} \min_{d(\cdot)} l(\zeta(0;z, t, u(\cdot), d(\cdot)))
\end{equation}
Note that the roles of $u(\cdot)$ and $d(\cdot)$ are opposite and reversed in the case
of minimal and maximal sets.
Because of the dynamic programming principle of optimality, the value function $\phi(z, s)$ satisfies the following Hamilton-Jacobi Partial Differential Equation (PDE):
\begin{equation}
    \label{eq: hj_pde}
    \begin{gathered}
        \frac{\partial\phi}{\partial s}(z, s) + \min_{d \in \mathcal D}\max_{u \in \mathcal U} \frac{\partial\phi}{\partial z}(z, s)^\top f(z, u, d) = 0\\
        \phi(z,0) = l(z), s \in [t, 0] \end{gathered}
\end{equation}
Thus by solving this PDE, we can obtain the function $\phi(z, s)$ and subsequently its
BRS by considering the sub-zero level set.
Our toolbox's implementation of the algorithm based on level-set methods for solving
equation \ref{eq: hj_pde} is illustrated in \textbf{algorithm} \ref{algo:Time-dependent
HJ PDE}. 
\begin{algorithm}
    \caption{Algorithm for solving HJ PDE}
    \label{algo:Time-dependent HJ PDE}
    \begin{algorithmic}[1]
        \State Initialize grid $g, \phi(z, s=0)$
        \State Initialize $t=0$, compute horizon $T$
        \State \textcolor{red}{// Hamiltonian computation}
        \While{$t \geq T$}
        \For{ every grid point index $i$ }
        \State \texttt{Compute $\frac{\partial\phi}{\partial z}^{+}(z_i, s)$,  $\frac{\partial\phi}{\partial z}^{-}(z_i, s)$}
        \Comment{\textcolor{red}{Forward and backward spatial derivative using ENO/WENO scheme}}
        \State  $\frac{\partial\phi}{\partial z}(z_i, s) \leftarrow \dfrac{1}{2}\left(\frac{\partial\phi}{\partial z}^{+} + \frac{\partial\phi}{\partial z}^{-}\right)$
        \State  $\displaystyle u_\text{opt}, d_\text{opt} \leftarrow \arg \min_{d \in \mathcal D} \max_{u \in \mathcal U} \frac{\partial\phi}{\partial z}(z_i, s)^{\top}f(z_i,u, d)$   \Comment{\textcolor{red}{Optimal control and disturbances based on user-defined objectives}}
        \State $H_{i} \leftarrow \frac{\partial\phi}{\partial z}(z_i, s)^{\top}f(z_i, u_\text{opt}, d_\text{opt})$
        \State $ \text{D}_i^{min} \leftarrow \min(\text{D}_i^{min},\frac{\partial\phi}{\partial z}(z_i, s))$
        \State $ \text{D}_i^{max} \leftarrow \max(\text{D}_i^{max},\frac{\partial\phi}{\partial z}(z_i, s))$
        \EndFor
        \State \textcolor{red}{// Artificial dissipation for stabilization }
        \For{every grid point index i}
        \State $\alpha_{i} \leftarrow \max_{p \in [\text{D}_i^{min}, \text{D}_i^{max}]} \abs{\dfrac{\partial H}{\partial p}(z_i, s)}$
        \Comment{\textcolor{red}{$H = p^\top f$}}
        \State $H_{i} \leftarrow H_{i}- \dfrac{1}{2}\alpha_{i}^{\top} \left( \frac{\partial\phi}{\partial z}^{+}(z_i, s) - \frac{\partial\phi}{\partial z}^{-}(z_i, s) \right) $ \Comment{\textcolor{red}{Lax-Friedrichs Scheme}}
        \State $\alpha^\text{max} \leftarrow \max(\alpha^\text{max}, \alpha_{i})$
        \EndFor
        \State \textcolor{red}{// Courant–Friedrichs–Lewy step}
        \State $\Delta s \leftarrow \left(\sum_{d=1}^{N}\dfrac{\alpha^\text{max}[d]}{\Delta z_{d}} \right)^{-1}$
        \State $\phi(z, s - \Delta s) \leftarrow H\Delta{s} + \phi(z, s)$  \Comment{\textcolor{red}{Runge-Kutta method for time integration}}
        \State $t = t - \Delta s$
        \EndWhile
    \end{algorithmic}
\end{algorithm}

\begin{figure}
    \centering
    \includegraphics[width=1.\textwidth]{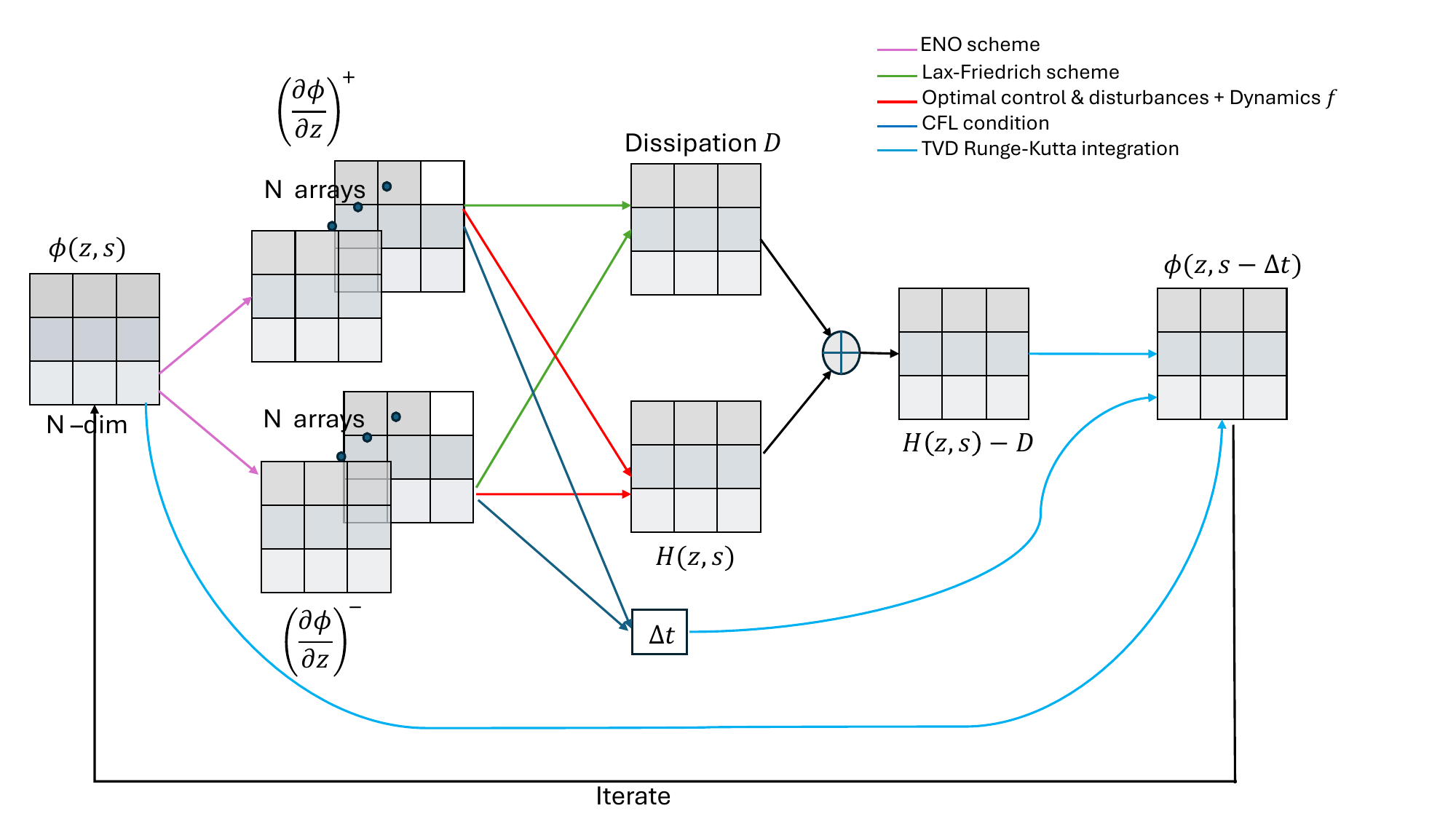}
    \caption{Illustration of stages in numerical process of solving time-dependent HJ PDE. 
    In ToolboxLS \cite{LsetToolbox1}, temporary variables are stored in multidimensional arrays as the same size of the grid.
        As we increase the number of dimensions, the DRAM memory required for these temporary
        array goes up linearly.
        If the depth of the computation is large, the total amount memory used for temporary
        variables will exceed system's DRAM capabilities, limiting computations to
        low-dimensional control problem only.}
    \label{fig:MATLAB_implementation}
\end{figure}

\begin{figure}[H]
    \centering
    \includegraphics[width=0.60\textwidth]{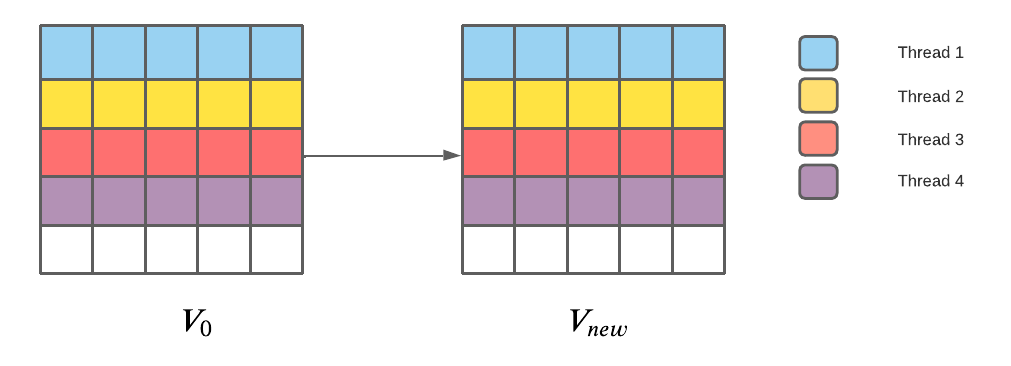}
    \caption{OptimizedDP's implementation of \textbf{algorithm \ref{algo:Time-dependent HJ PDE}}  does not buffer temporary variables into multidimensional arrays.
        Instead, within each grid iteration, a grid point value in $V_{\text{new}}$ is directly
        computed. Each thread is assigned a chunk of grid points for parallel computation.
    }
    \label{fig:our_implementation}
\end{figure}

Notice that our implementation of our \textbf{algorithm} \ref{algo:Time-dependent HJ PDE}
is based on imperative programming, where we specify the computation procedure for each grid point.
Alternatively, one can consider a vectorized approach that computes, stores, and operates on each component 
of the algorithm for the entire grid at once, which is implemented in ToolboxLS and hj\_reachability
\cite{hjreachability,LsetToolbox1}. This vectorized approach is illustrated in Figure \ref{fig:MATLAB_implementation}, which provides a graphical overview of
the stages and components
involved in the numerical process of solving time-dependent HJ PDEs.
Although this approach supports an arbitrary number of dimensions through the
usage of various operation tricks, it comes with the cost of extra memory usage.
In particular, the temporary variables
such as spatial derivatives, system dynamics, etc. for the whole grid is stored in multiple
grid-sized arrays.
This approach is not ideal for the performance of an already expensive computation in
two ways (illustrated in \textbf{Fig.} \ref{fig:MATLAB_implementation}).
Firstly, the approach does introduce extra overhead of memory in the implementation.
These redundant overheads increase linearly as we go up the dimensional ladder, which
can limit the number of dimensions to which the algorithm can be performed.
Secondly, each of the components for all grid points has to be computed before the
final output, which results in bad cache locality for high-dimensional problems. On the other hand,
algorithm \ref{algo:Time-dependent HJ PDE} does not buffer temporary variables into multidimensional arrays, but 
directly maps each grid point value to a new value in $V_{\text{new}}$ as illustrated in Figure \ref{fig:our_implementation}.

In \textbf{Algorithm} \ref{algo:Time-dependent HJ PDE}, it is often possible to eliminate the second for loop that computes the stabilizing artificial dissipation by approximating
the term $\alpha_{i}$ as the largest of rates of changes over the defind grid $g$ for each component \cite{Osher2002LevelSM}. 
This can help halve the computational time and reducing memory consumption at the cost of approximation error of the 
numerical solution.

Additionally, our toolbox allows solving variations of equation 
\eqref{eq: hj_pde} such as the HJ variational inequality, which is essential for computing the \textbf{BRT}:
\begin{equation}
    \label{eq:hji_variational_inequality}
    \begin{aligned}
        \min \left\{ \frac{\partial\phi}{\partial s}(z, s) + \min_{d \in \mathcal D}\max_{u \in \mathcal U} \frac{\partial\phi}{\partial z}(z, s)^\top f(z, u, d), l(z)-\phi(z, s) \right\} & =0, \quad s \in[t, 0] \\
        \phi(z,  t )                                                                                                                                                                        & = l(z)
    \end{aligned}
\end{equation}

Our toolbox also supports solving other variations of the above equation developed for
time-varying target set and reach-avoid games formulation developed in
\cite{Fisac2014ReachavoidPW}.
It lets users choose to either compute reachable Set or Tubes as well as specifying
time-varying obstacle set and target set, which involves solving the corresponding PDE
specified by users.
Currently, the toolbox can work with dynamical systems with up to 6-8 dimensions,
depending on the available hardware resources and stiffness of the dynamical system.

\subsection{Time-independent (static) Hamilton-Jacobi (HJ) Partial Differential Equation (PDE) }
In addition, optimizedDP provides an implementation of the Lax-Friedrich sweeping
algorithm based on the work of \cite{TTR} for efficiently computing the reachable set without time integration. At its core, the algorithm iterate through the grid and update each point using Gauss-Siedel iteration method
where the update formula is given by the Lax-Friedrichs scheme.

Given a target set $\mathcal{T} \subseteq  \mathcal{R}^n$, the time-to-reach (TTR) function is defined as follows:
\begin{equation}
    \phi(z) = \max_{d(\cdot)\in \mathbb D} \min_{u(\cdot) \in \mathbb{U}} \min\{t \mid z(t) \in  \mathcal{T}\}
    \label{ttr_definition}
\end{equation}

By dynamic programming principle, this TTR function $\phi(z)$ can be obtained by solving the following HJ PDE \cite{Isaac1965DifferentialGames}:
\begin{equation}
    \label{eq:TTR_HJ_variational_inequality}
    \begin{gathered}
        H\left(z, \frac{\partial\phi}{\partial z}(z)\right) = 0\\
        \phi(z) = 0, z \in \mathcal{T} \\
        H\left(z, \frac{\partial\phi}{\partial z}(z)\right) =\min_{u\in \mathcal U} \max_{d\in \mathcal D} \left( -\frac{\partial\phi}{\partial z}(z)^{\top} f(z, u, d) - 1\right) \end{gathered}
\end{equation}
To solve for $\phi(z)$, \textbf{Algorithm} \ref{algo: lax_friedrich sweeping} proposed in \cite{TTR} can be used for efficient computation.
Compared to solving the time-dependent HJ PDE, \textbf{algorithm} \ref{algo: lax_friedrich sweeping} requires less memory and the convergent result generally
requires fewer iterations. In the beginning of this algorithm, we initialize $\phi(z)$ to be zero for all grid points in the target 
set $\mathcal{T}$ and infinity for all other grid points. In our implementation, since infinity is not valid value, we assign these 
points to a large value such as $10000$. 
Then, we iteratively update each grid point using the Lax-Friedrichs scheme until convergence.

\begin{algorithm}
    \caption{Lax-Friedrichs sweeping algorithm \cite{TTR}}
    \label{algo: lax_friedrich sweeping}
    \begin{algorithmic}[1]

        \State Initialize $\phi(z) \leftarrow 0$ for $z \in \mathcal{T}$ and $\phi(z) \leftarrow \infty$ for $z \not\in \mathcal{T}$
        \While {$\abs{\phi - \phi^\text{old}} < \epsilon$}
        \State $\phi\leftarrow \phi^\text{old}$
        \For{ grid index $i$ not in boundary}:
        \State \texttt{Compute $\frac{\partial\phi}{\partial z} (z, s)$}
        \State  $\displaystyle u_\text{opt} \leftarrow \arg \min_{u \in \mathcal{U}} \frac{\partial\phi}{\partial z}(z, s)^{\top}f(z,u)$
        \State $\dot{z} \leftarrow f(z, u_\text{opt})$
        \State  $H_{i} \leftarrow \frac{\partial\phi}{\partial z}(z, s)^{\top}\dot{z}$
        \State $\sigma \leftarrow \abs{\dfrac{\partial H}{\partial p}}$
        \State $c \leftarrow \frac{\Delta z}{\sigma}$
        \State $\phi_{i}^\text{new} \leftarrow c(-H_{i} + \sigma\frac{\phi_{i+1} + \phi_{i-1}}{2\Delta z} )$
        \State $\phi_{i} \leftarrow \min(\phi^\text{new}_{i}, \phi_{i})$
        \EndFor
        \State // Update the grid points at boundary
        \State $\phi^\text{new}_{1} \leftarrow \min(\max(2\phi_{2} - \phi_{3}, \phi_{3}), \phi_{1} ) $
        \State $\phi^\text{new}_{N} \leftarrow \min(\max(2\phi_{N-1} - \phi_{N-2}, \phi_{N-2}), \phi_{N} ) $
        \EndWhile

    \end{algorithmic}
\end{algorithm}

\subsection{ Discretized Value Iteration for Markov Decision Process (MDP)}
Markov Decision Process is a useful model for studying the optimal behavior of a
target system in reaction to the changes in external environments.
An MDP is usually described by a tuple $(S, A, T, R, \gamma, H)$ where $S$ is the state
space, $A$ is the action space, $T$ is the transition probability matrix, $\gamma$ is
the discount factor, $R$ is the reward signal, and $H$ is the time horizon.
The key assumption of MDP is the next state transition of a system is only dependent on
the current state and action.
This assumption is described by the following relation

\begin{equation}
    \mathbf{P}(s_{t+1} | s_{t}, a_{t}) = \mathbf{P}(s_{t+1} | s_{t},a_{t} ..., s_{0}, a_{0})
    \label{ttr_definition}
\end{equation}

\noindent where $s_{t} \in S$, and $a_{t} \in A$.
In MDP, the discounted return $G_{t}$ at time step $t$ is defined as

\begin{equation}
    G_{t} = R_{t+1} + \gamma R_{t+2} + \gamma^{2} R_{t+3} + .
    .. = \sum^{n}_{k=0}\gamma^{k} R_{t+k},
    \label{ttr_definition}
\end{equation}

\noindent and the state value function $V_{\pi}(s)$ for $s \in S$ under a policy $\pi: S \rightarrow A$ is as
\begin{equation}
    V_{\pi}(s) = E_{\pi}[G_{t}| S_{t} = s] = E_{\pi}[R_{t} + \gamma V_{\pi}(s') | S_{t} = s].
    \label{ttr_definition}
\end{equation}

In an MDP, the objective of the target system is to act according to an optimal policy $
    \pi^{*} : S \rightarrow A $ that can maximize the expected rewards received at each
state over time. Our goal in MDP is to compute $\pi ^ {*} $ along with the maximum expected rewards
received at every state:
\begin{equation}
    \max_{\pi} E_{\pi}[G_{t}| S_{t} = s] = \max_{\pi} V_{\pi}(s)
    \label{RL_objective}
\end{equation}

This objective and the basic properties of MDP are the backbone of all reinforcement
learning algorithms. Our toolbox provides an implementation of the value iteration algorithm in
\cite{sutton2018reinforcement} for continuous state and action space (shown in
\textbf{Algorithm} \ref{Value Iteration}), which computes expected rewards
$V_{\pi^{*}}(s)$ at every state given all the possible actions a state $s$ can take.
Note that at line 8 of algorithm \ref{Value Iteration}, $(s')$ is obtained by
considering the nearest neighbor that is the closest discretized state on the grid based on dynamics/
transition model.

\begin{algorithm}
    \caption{Value Iteration Algorithm - Continuous MDP}
    \label{Value Iteration}
    \begin{algorithmic}[1]
        \State Discretize $S, A$
        \State $V_{t=0} \leftarrow 0$
        \State $\Delta \leftarrow 0$
        \State Repeat:
        \For{ $s$ in $S$}
        \For{ $a$ in $A$}
        \State $v \leftarrow V(s)$
        \State $Q(s, a) \leftarrow R(s, a) + \sum_{s'} p(s' | s, a) V(s')  $
        \State $V(s) \leftarrow \max (V(s), Q(s, a) )  $
        \State $ \Delta \leftarrow \abs{V(s) - v}$
        \State If $\Delta > $ threshold:
        \State\hspace{\algorithmicindent} Repeat next iteration
        \EndFor
        \EndFor
    \end{algorithmic}
\end{algorithm}

\section{Overview of the Toolbox Structure}

\begin{figure*}[h]
\includegraphics[width=0.9\textwidth]{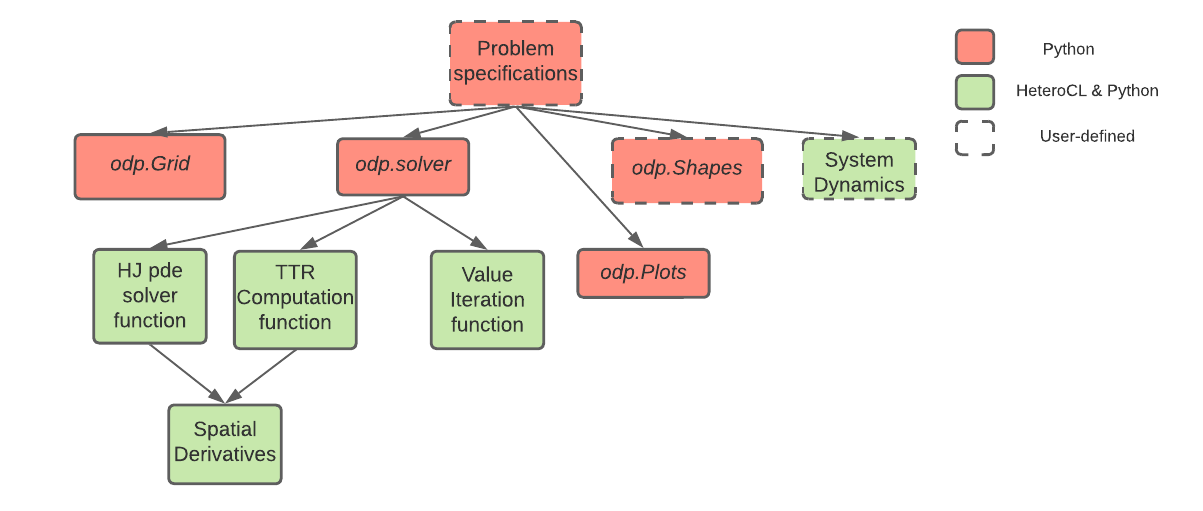}
\caption{The overall structure of OptimizedDP consists of red blocks
 (Python with NumPy) for problem specification, grid initialization,
and plotting, and solid green blocks (Python/HeteroCL) for core algorithms. 
User-specified system dynamics object containing problem parameters and 
subroutines for optimal controls are then plugged into core solvers.}

\label{fig:toolbox_structure}
\end{figure*}

The general structure of our toolbox is shown in \textbf{Figure}
\ref{fig:toolbox_structure}.
Our software library provides different agorithm implementations, solver function calls to access
these implementations, a set of libraries to numerically initialize the problem 
and visualizing utilities functions to display the results. The core interface of the toolbox resides
in the \textit{odp.solver} module, which provides solver function calls to access different algorithm implementations. 
When a solver function in \textit{odp.solver} is called, it will in turn call the corresponding 
algorithm implementation written in HeteroCL which builds, optimize a computational graph and
return a function-like executables. 
In \textit{odp.solver}, these executables are then called iteratively as new input arguments are passed to
and process the output based on computation modes. 
Additionally, the core algorithm implementations in HeteroCL all allows plug-in system dynamics modules.
These system dynamics modules are user-defined Python objects that 
contains problem parameters and subroutines to compute specific dynamical components of the target algorithm
such as optimal controls, rate of change for each system state, transition matrix, etc.

Additionally, to initialize the numerical grid and boundary conditions, our toolbox provides extendable
libraries which include Cartersian grid generation (package \textit{odp.Grid}), and
initialization of signed distance function for different shapes (package
\textit{odp.Shapes}). These packages are all written in Python and Numpy libraries,
which could be easily extended and customized by users. 
Once the results are computed and converted to a Numpy array, available visualization libraries in the
toolbox can be used to display the result.
To make visualization of high dimension array easier, the package
\textit{odp.Plots} provides API functions that can be called to visualize 3D isosurface and 2D contours of
the value function with options to index the multidimensional result array.
A more detailed discussion of the features of our toolbox and how to use the toolbox with concrete examples will be discussed in more
details in the next few sections.

\subsection{Library Components and Features}

\subsubsection{Grid}
Similar to the ToolboxLS \cite{LsetToolbox1}, our toolbox allows users to create a
Cartesian grid, implemented as a Python object, by specifying the number of grid nodes,
upper bound, lower bound for each dimension, and periodic dimension.
The ghost points at the boundary for the non-periodic dimension, by default, are
extrapolated to maintain the same sign of the boundary points. The implementation
of the grid structure can be found in the \textit{odp.Grid} module (Fig. \ref{fig:toolbox_structure}).

\subsubsection{Initial Condition}
To initialize different implicit surface shapes, we have implemented many initial
conditions that represent shapes such as cylinders, spheres, and lower/upper planes.
In addition, there are utility functions that operate on these geometry shapes such as
union and intersection.
All of these functions are written with Python and Numpy, and could be easily extended
by users using the attribute \textit{grid.vs} exposed by the \textit{grid} object.
The implementation of these initial conditions can be found in
the \textit{odp.Shapes} module (Fig. \ref{fig:toolbox_structure}).

\subsubsection{Time Integration}
OptimizedDP provides implementations of first-order and second-order accurate total
variation diminishing (TVD) Runge-Kutta (RK) integration methods for solving the
dynamic HJ PDE.
The maximum timestep used for integration is determined by the Courant–Friedrichs–Lewy
(CFL) \cite{CFL} condition. The time integration methods are implemented in the
\textit{odp.solver} module.

\subsubsection{Spatial Derivatives}
Currently, OptimizedDP provides an implementation of the derivatives approximation
method that includes first-order upwind approximation and second-order accurate
essentially non-oscillatory (ENO) \cite{ENO} \cite{Osher2002LevelSM} scheme. The implementations
of these methods can be found in the \textit{odp.spatialDerivatives} module.
Higher-order schemes such as third-order ENO and weighted ENO schemes will be added to
later version releases of our toolbox.

\subsubsection{Visualization}
OptimizedDP provides two options for visualizing computational results. Both allow users to 
visualize low-dimensional sublevel sets of the high-dimensional value function. These functions
can be found in the package \textit{odp.Plots} (Fig. \ref{fig:toolbox_structure}).
\begin{enumerate}
    \item \textit{plot\_isosurface}: This function visualize 3D or 2D sublevel set of a an input value function.
          At its core, the interface utilizes \textit{plotly} library's \textit{Isosurface} function for 3D and
          \textit{Contour} for 2D function visualization in a browser.
    \item \textit{plot\_valuefunction}: This interface allows to visualize 2D or 1D value function results.
          The interface calls the function \textit{Surface} for 2D and \textit{Scatter} for 1D
          available in \textit{plotly} library, which will show the value at different grid
          points and also highlight the zero sublevel set in the visualization result.
\end{enumerate}

\begin{figure}[h]
    \centering
    \includegraphics[width=1.\linewidth]{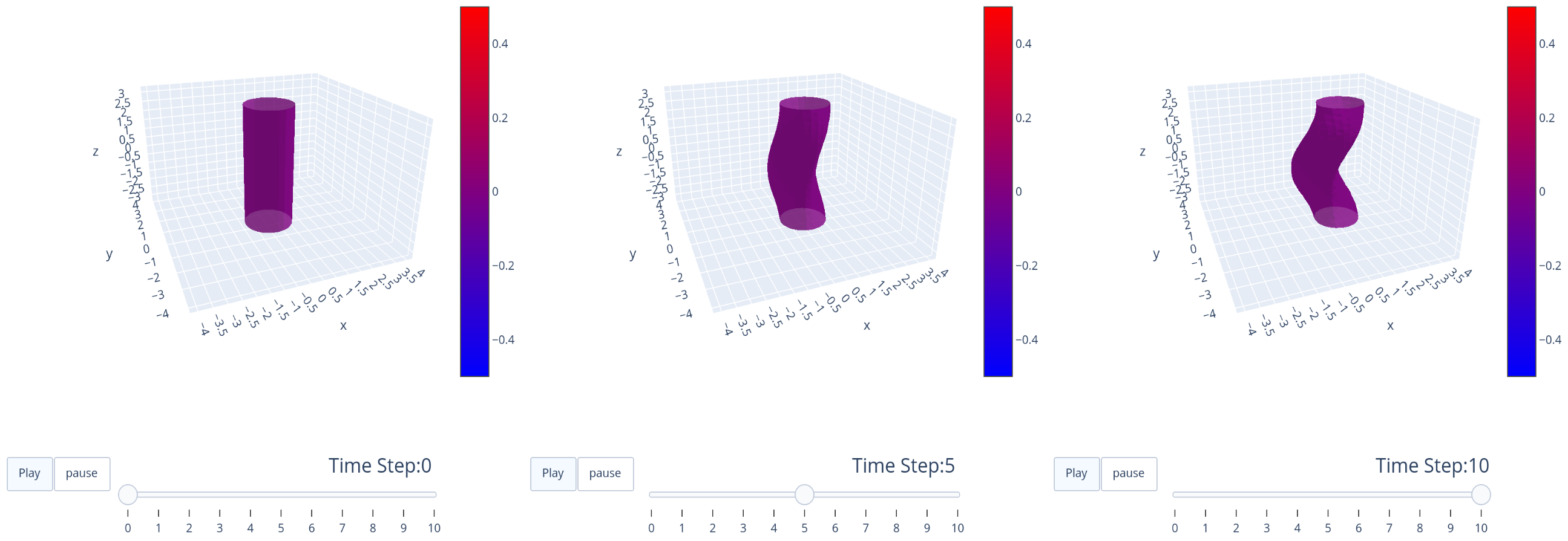}
    \caption{3D visualization of sub-zero level set across different timestep. User can choose value function at certain timestep to visualize}
    \label{fig:enter-label}
\end{figure}
\section{Coding Example}
In this section, we showcase an example to demonstrate the ease of specifying a problem
of interest with minimal programming effort, while still benefitting from the much
faster computation times compared to similar toolboxes.
All the examples discussed in this section can be found at the GitHub repository
\url{https://github.com/SFU-MARS/optimized_dp/examples}.

\subsection{Initializations}
When solving any one of the supported algorithms, at the beginning, users first need to specify 
the grid object over which the PDEs are solved by specifying its bounds and the number of grid points in each dimension.
Using this grid instance, the initial value function of the PDE can be then generated by calling the utility functions
from library packages odp.Grid and odp.Shapes. An example of this is shown in the code snippet below.

\begin{lstlisting}[language=Python, numbers=left, escapechar=|]
    import numpy as np
    from odp.Grid import Grid
    from odp.Shapes import *
    from dynamics.DubinsCapture import *
    from plot_options import *
    from solver import HJSolver
    
    import math
    
    # Reach-Avoid Example
    g = Grid(np.array([-4.0, -4.0, -math.pi]), np.array([4.0, 4.0, |\label{code:grid}|
            math.pi]), 3, np.array([40, 40, 40]), [2])
    
    # Reachable set
    target_set = CylinderShape(grid=g, ignore_dims=[2], 
                         center=np.zeros(3), 
                         radius=0.5) |\label{code:target}|
\end{lstlisting} 

In the above, we have initialized a grid of size $40$ x $40$ x $40$ over the states range of $x \in [-4, 4]$, 
$y \in [-4., 4.]$, $\theta \in [-\pi, \pi]$. The \lstinline|CylinderShape| function takes in this grid, and at each grid point, it computes the signed distance 
with respect to a cylinder surface of radius $0.5$ centered at the origin, which basically is $\phi(z, 0) = \sqrt{x^2 + y^2} - 0.5$.
The argument \lstinline|ignore_dims| specifies which dimensions of the grid to ignore when computing this function. In this case, we ignore 
the third dimension (second-indexed), which is the angle $\theta$. For value iteration computation, similarly a grid with bounds 
and number of grid points in each dimension is needed to be specified. On the other hand, the value function needs not to be specified
at the beginning.

\subsection{Dynamical Systems Specification}
Our example in this section illustates a coding example of the dynamics for Pursuit-Evasion game between two Dubins car systems \cite{MerzIdenticalCars}.
The reduced order model of the game is the relative states between the two players whose 
evolution are described by the following set of differential equations:

\begin{equation}
    \label{eq:pursuit_evasion_dynamics}
    \begin{aligned}
        \dot{x}    & = -v + v\cos\theta + ay\\
        \dot{y}   & = v \sin\theta - ax           \\
        \dot{\theta} & = b-a
    \end{aligned}
\end{equation}
where $\abs{a} \leq A$,  $\abs{b} \leq B$ are the angular control inputs of the evader and pursuer respectively, while $v$ is the constant speed of both evader
and pursuer. In this game, the evader is trying to run away from the pursuer by maximizing the relative distance while the pursuer is 
trying to minimize this value. In this case, the implicit capture target function can be written as $\phi(x, y, \theta) = \sqrt{x^2 + y^2} - R$,
where $R$ is the capture radius. Next, we can expand the Hamiltonian term of the HJ PDE as follows:
\begin{equation}
    \label{eq:pursuit_evasion_hamiltonian}
    H =  \max_{a} \min_{b} \left[\dfrac{\partial \phi}{\partial x}(-v_a + v_b\cos\theta + ay)  + 
                 \dfrac{\partial \phi}{\partial y}(v_a\sin\theta - ax) +
                 \dfrac{\partial \phi}{\partial \theta}(b-a) \right]
\end{equation}

Since the evader is maximizing and the pursuer is minimizing, the optimal control and disturbance
are can be compactly written as $u_{\text{opt}} = a_{\text{opt}} =  \text{sign} \left(\dfrac{\partial \phi}{\partial x} y - \dfrac{\partial \phi}{\partial y}x - \dfrac{\partial \phi}{\partial \theta} \right) A$
and $d_{\text{opt}} = b_{\text{opt}} =  -\text{sign} \left(\dfrac{\partial \phi}{\partial \theta} \right) B$ respectively.

To compute the winning regions of each player or the time to capture, we can respectively solve the time-dependent and time-independent HJ PDE for the above
dynamics.
In order to do so, user need to provide a system dynamics object that must contain three subroutines \lstinline{opt_ctrl}, \lstinline{opt_dstb}, and \lstinline{dynamics}. 
The functions \lstinline{opt_ctrl} and \lstinline{opt_dstb} contain the logic to determine the optimal control $a$
and optimal disturbance $b$ as described. These functions return a fixed-size tuple of control and disturbances inputs respectively,
to compute the Hamiltonian term \eqref{eq:pursuit_evasion_hamiltonian}. Users can also include static physical parameters of the systems in the class constructor \lstinline{__init__} that 
can be used inside the object's member functions. The following code snippet of \lstinline{DubinsCapture} class illustates how to write these functions.

\begin{lstlisting}[language=Python]
import heterocl as hcl

class DubinsCapture:
    def __init__(self, x=[0,0,0], wMax=1.0, speed=1.0, dMax=1.0,
            uMode="max", dMode="min"):
        self.x = x
        self.wMax = wMax
        self.speed = speed
        self.dMax = dMax
        self.uMode = uMode
        self.dMode = dMode

    def opt_ctrl(self, t, state, spat_deriv):

        opt_w = hcl.scalar(0, "opt_w")
        
        # Declare a variable
        a_term = hcl.scalar(0, "a_term")
        # use the scalar by indexing 0 everytime
        a_term[0] = spat_deriv[0] * state[1] - spat_deriv[1] * state[0] 
                                                        - spat_deriv[2]
        # Python condition for static variable
        if self.uMode == "max":
            # HeteroCL condition for runtime variable
            with hcl.if_(a_term >= 0):
                opt_w[0] = self.wMax
            with hcl.elif_(a_term < 0):
                opt_w[0] = -self.wMax
        # Dummy values to be returned
        in3 = hcl.scalar(0, "in3")
        in4 = hcl.scalar(0, "in4")

        return (opt_w[0], in3[0], in4[0])

    def opt_dstb(self, t, state, spat_deriv):

        d1 = hcl.scalar(0, "d1")
        
        # Python condition for static variable
        if self.dMode == "min":
            # HeteroCL condition for runtime variable
            with hcl.if_(spat_deriv[2] >= 0):
                d1[0] = -self.dMax
            with hcl.elif_(spat_deriv[2] < 0):
                d1[0] = self.dMax

        # Dummy values to be returned
        d2 = hcl.scalar(0, "d2")
        d3 = hcl.scalar(0, "d3")
        return (d1[0], d2[0], d3[0])

    \end{lstlisting}

In this particular example, the arguments \lstinline{state}, and \lstinline{spat_deriv} to the functions \lstinline|opt_ctrl|
and \lstinline|opt_dstb| corresponds to the state vector $(x, y, \theta)$ and spatial derivative vector 
$\left(\dfrac{\partial \phi}{\partial x}, \dfrac{\partial \phi}{\partial y}, \dfrac{\partial \phi}{\partial \theta}\right)$ respectively.
The argument \lstinline|t| is not part of the computations anywhere in this example, but it is included in the function signature
that generalizes for time-varying system dynamics. 
Once the optimal control and disturbances have been determined, they are then passed in to a dynamics function that computes and 
returns a tuple of rate of changes of each state component, which is $f(z_i, u, d)$
used in Algorithm \ref{algo:Time-dependent HJ PDE}.

\begin{lstlisting}[language=Python]
    def dynamics(self, t, state, uOpt, dOpt):
        x_dot = hcl.scalar(0, "x_dot")
        y_dot = hcl.scalar(0, "y_dot")
        theta_dot = hcl.scalar(0, "theta_dot")

        x_dot[0] = -self.speed + self.speed*hcl.cos(state[2]) + uOpt[0]*state[1]
        y_dot[0] = self.speed*hcl.sin(state[2]) - uOpt[0]*state[0]
        theta_dot[0] = dOpt[0] - uOpt[0]

        return (x_dot[0], y_dot[0], theta_dot[0])
\end{lstlisting}
All three functions \lstinline{opt_ctrl},  \lstinline{opt_dstb}, and \lstinline{dynamics} are written in Heterocl
instead of pure Python functions, where each variable is declared as a \lstinline{hcl.scalar} and the logic statements are \lstinline{hcl.if_} and
\lstinline{hcl.elif_}. The reason is these functions are plug-in modules to the core algorithm implementation written in Heterocl
that are optimized at compiled time for performance. 

In the next example, our system is an inverted pendulum system and our goal is to balance the pendulum at the upright position
by applying a torque to one end of the pendulum. This is a classical control problem for reinforcement learning.
The dynamics and rewards function of the system are as follow:

\begin{equation}
    \label{eq:pendulum_dynamics}
    \begin{aligned}
        \dot{\theta}  & = \omega \\
        \dot{\omega}  & = \left(\dfrac{3 g}{2 l} \right) \omega + \left(\dfrac{3}{m l^2} \right) u \\
    \end{aligned}
\end{equation}
where $l$ is the length of the pendulum, $m$ is the mass of the pendulum, $g$ is the gravitational constant, 
and $u$ is the torque applied to the pendulum. All of these parameters are specified in the constructor of the class.
And the reward function is

\begin{equation}
    \label{eq:car_reward}
    r(s, u) = -\left( \theta^2 + 0.1 \omega^2 + 0.001 u^2 \right)
\end{equation}

Any systems input to the value iteration solver need to have a transition \lstinline|transition| function which returns the next states  
with their probabilities matrix and the \lstinline|reward| function returns the reward given a state and action. The code
snippet below shows the implementation of the transition and reward functions for the inverted pendulum system described.
\begin{lstlisting}[language=Python]

   class pendulum_2d_example:
    def __init__(self):
        # Some constant parameters for pendulum adapted from the openAI gym
        self.dt = 0.05
        self.g = 10
        self.m = 1.
        self.l = 1.
        self.max_speed = 8.
        self.coeff1 = 3 * self.g/ (2* self.l)
        self.coeff2 = 3.0/(self.m * self.l * self.l)
        self.maxTransitions = 1 

    def transition(self, sVals, iVals, u):
        trans_matrix = hcl.compute((self.maxTransitions, (1 + 2)),
                        lambda *x: 0, "trans_matrix")
        # Variable declaration
        newthdot = hcl.scalar(0, "newthdot")
        th = hcl.scalar(0, "th")
        new_th = hcl.scalar(0, "new_th")
        # Just use theta from goals variable
        th[0] = sVals[0]

        newthdot[0] = sVals[1] + (self.coeff1 * hcl.sin(sVals[0]) 
                                    +  self.coeff2 * u) * self.dt
        # Probability of 1 for deterministic transition 
        trans_matrix[0, 0] = 1.0
        trans_matrix[0, 1] = new_th[0]
        trans_matrix[0, 2] = newthdot[0]

        return trans_matrix

    # Return the reward for taking action from state
    def reward(self, sVals, iVals, u):
        # Variable declaration
        rwd = hcl.scalar(0, "rwd")
        rwd[0] = -(sVals[0] * sVals[0] + 0.1 * sVals[1] * sVals[1] + 0.001 * u * u)
        return rwd[0]

\end{lstlisting}

\subsection{Solver Initiation}
\begin{table}[h!]
\centering
\begin{tabular}{|c|c|c|c|}
\hline
\textbf{Method} & \textbf{Description} & \textbf{Operation}  \\ \hline
minVWithV0 & Minimum with Initial Value & $\phi_{t+1} = $ min$(\phi_{t+1}, \phi_{0})$ \\ \hline
maxVWithV0 & Maximum with Initial Value & $\phi_{t+1} = $ max$(\phi_{t+1}, \phi_{0})$ \\ \hline
minVWithVInit &  Minimum Value Over Time & $\phi_{t+1} = $ min$(\phi_{t+1}, \phi_{t-1})$ \\ \hline
maxVWithVInit & Maximum Value Over Time &  $\phi_{t+1} = $ max$(\phi_{t+1}, \phi_{t-1})$  \\ \hline
minVWithVTarget & Minimize Value with Target Set & $\phi_{t+1} = $ min$(\phi_{t+1}, l_{t-1})$ \\ \hline
maxVWithVTarget & Maximize Value with Target Set & $\phi_{t+1} = $ max$(\phi_{t+1}, l_{t-1})$ \\ \hline
minVWithObstacle & Minimize Value with Obstacle Set &  $\phi_{t+1} = $ min$(\phi_{t+1}, g_{t-1})$ \\ \hline
maxVWithObstacle & Maximize Value with Obstacle Set &  $\phi_{t+1} = $ max$(\phi_{t+1}, g_{t-1})$ \\ \hline
\end{tabular}
\caption{Computation Methods for Value Function}
\label{tab:computation_methods}
\end{table}

After intitializing the grid, initial value function and specifiying the system dynamics, users can now call
the core solver functions of interest.
For solving the time-dependent HJ PDE,  the target solver function is \lstinline|HJSolver|. When 
calling the function, integration time and time increments at which the value function is integrated are
passed to the function. Additionally, there are certain computation methods that can be specified to compute 
the value function. All the computation methods summarized in Table \ref{tab:computation_methods}. Depending on the computation method specified, different versions of the HJ pde is solved. For example, if the method is set to 
be \lstinline|"None"|, the solver will compute the backward reachable set \ref{eq:brs} by solving Eq. \eqref{eq: hj_pde}. On the other hand,
if the method is set to \lstinline|"minVWithV0"|, the solver will compute the backward reachable tube \ref{eq:brt} 
by solving Eq. \eqref{eq:hji_variational_inequality}. If one wishes to solve a time-varying reach-avoid problem, the computation method
can be set to \lstinline|"minVWithVTarget"| and \lstinline|"maxVWithObstacle"|, with a list of the target and obstacle sets
 $[l(z), g(z)]$ then passed to the \lstinline|HJSolver| function which will then solve
the PDE formulation in \cite{Fisac2014ReachavoidPW}. Depending on the problem, user can also choose to return the value function
at all time step by setting the \lstinline|saveAllTimeSteps| argument to \lstinline|True|.
An example of doing this is shown in the code snippet below:

\begin{lstlisting}[language=Python, numbers=left, escapechar=|]
    # Look-back length and time step
    lookback_length = 1.5
    t_step = 0.05
    small_number = 1e-5
    tau = np.arange(start=0, stop=lookback_length + small_number,
            step=t_step)
    my_car = DubinsCapture(uMode="min", dMode="max") |\label{code:dyn}|
    compMethods = { "TargetSetMode": "minVWithVTarget",
                    "ObstacleSetMode": "maxVWithObstacle"}
    result = HJSolver(my_car, |\label{code:solve}| g, [goal, obstacle], tau,
                        compMethods, saveAllTimeSteps=True )
\end{lstlisting}

Solving time-independent HJ PDE is also very similar to solving the time-dependent PDE. users can specify the time horizon and the time increments at 
which the value function is to be computed. The solver function to be called is \lstinline|TTRSolver|.

\begin{lstlisting}[language=Python, numbers=left, escapechar=|]
    my_car = DubinsCar(uMode="min", dMode="max") |\label{code:dyn}|

    epsilon = 1e-5
    result = V_0 = TTRSolver(my_car, g, targetSet, epsilon)
\end{lstlisting}

To compute value iterations, users call function  \lstinline|solveValueIteration| and pass in the defined pendulum system,
the grid, the action space, and other parameters of computation including the discount factor $\gamma$, the convergence threshold $\epsilon$,
and the maximum number of iterations \lstinline|maxIters|:

\begin{lstlisting}[language=Python, numbers=left, escapechar=|]
     result = solveValueIteration(pendulum_system,
                              grid=g, action_space=np.linspace(-2., 2., 41),
                              gamma=0.9, epsilon=1e-3,
                              maxIters=maxIters
                              )
\end{lstlisting}
The results returned from all three solver functions are Numpy array values that are ready to be visualized.

\subsection{Visualizing Outputs}
\begin{figure}
    \centering
    \begin{subfigure}[t]{.4\textwidth}
        \centering
        \includegraphics[width=\linewidth]{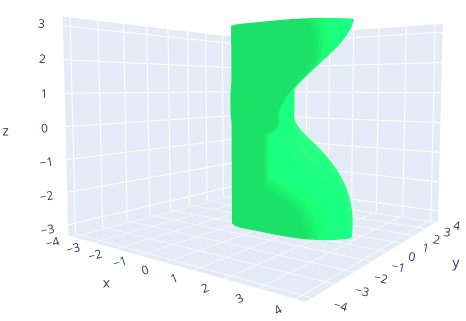}
        \caption{Left side view}
    \end{subfigure}
    \begin{subfigure}[t]{.4\textwidth}
        \centering
        \includegraphics[width=\linewidth]{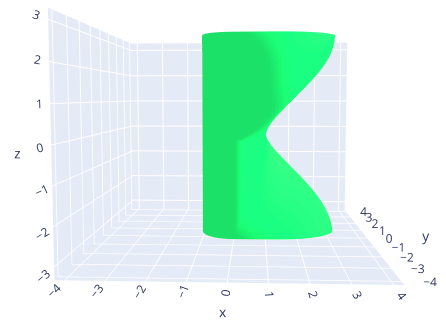}
        \caption{Right side view}
    \end{subfigure}
    \caption{ BRT or the sub-zero level set isosurface is plotted when 
    setting \lstinline|plot_type| to \lstinline|"set"| }
    \label{fig: DunbinsBRS}
\end{figure}

After obtaining the results, we can then visualize different slice of the high-dimensional value function,
using the prodvided visualizing function \lstinline|visualize_plots|.
When calling \lstinline|visualize_plots|, users need to pass in a plotting option object that specifies information of the plots
such as the plotting type (contour or value type) \lstinline|plot_type|, the list of dimensions to be fully plotted \lstinline|plotDims|, 
saving option \lstinline|save_fig| and
the indices of the missing dimensions \lstinline|slicesCut| over which the value function is 
indexed for plotting. After calling this function, the sub-zero $3$D surface plot of the value function in 
Pursuit-Evasion example is shown in Fig. \ref{fig: DunbinsBRS}.

\begin{figure}[H]
    \includegraphics[width=1\textwidth]{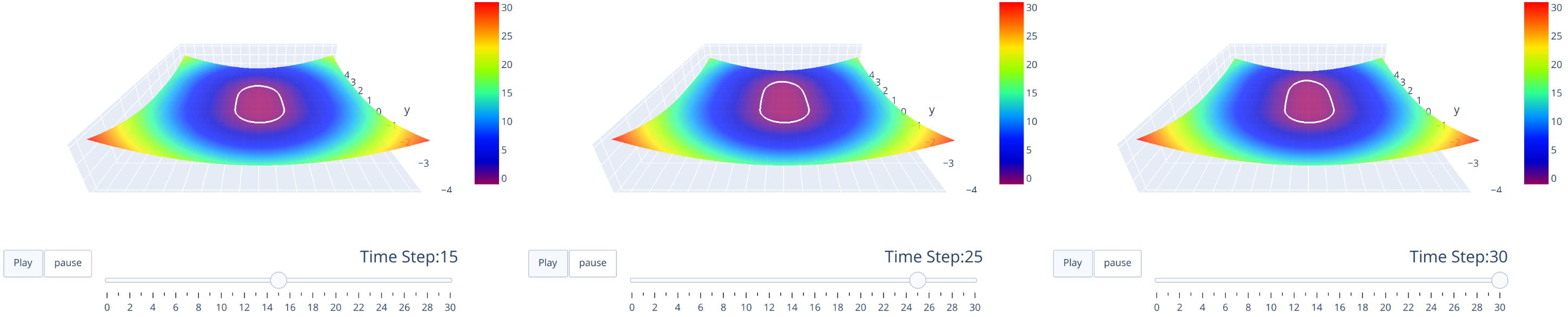}
    \caption{Instead of only visualizing a particular level set, user can choose to visualize the value function
    over the state domain by setting \lstinline|plot_type| to \lstinline|"value"|.
    The white contour in the plot illustrates the sub-zero level of this fuction}
    \label{fig:sliderbar}
\end{figure}

By varying the \lstinline|slicesCut| parameter and \lstinline|plotDims| parameter to the PlotOptions object,
users can plot and visualize different slices of the high-dimensional value function.
Additionally, if the input value function contains multiple time steps, users can also visualize 
how the value function evolves over time using a time slider (shown in Fig. \ref{fig:sliderbar}).
If the number of grids is too large to be visualized, the data will be automatically
downsampled for plotting efficiency.
A more detailed description of the plotting options can be found in the \lstinline|plot_options.py| file in our GitHub repository.
A snipped codes below shows how to specify the plotting options and visualize the results.

\begin{lstlisting}[language=Python]
    from plot_options import *
    from odp.utils import plot_3D_slice, plot_2D_slice

    grid = ... # The grid object created earlier
    results = ... # The result returned from the solver function

    # Plot isosurface of the value function
    po = PlotOptions(do_plot=True, plot_type="set", plotDims=[0,1,2],
                  slicesCut=[], save_fig=True, filename="test1.png")
    
    visualize_plots(results, grid, po) 
    po = PlotOptions(do_plot=True, plot_type="value", plotDims=[0,1],
                  slicesCut=[2], save_fig=True, filename="test2.png")
    # Plot the value function
    visualize_plots(results, grid, po)

\end{lstlisting}

\section{Optimization of Implementation}

In this section, we are going to discuss in more detail the optimization techniques
used to result in fast computation time.
These details of optimization are hidden from the user, who can focus on solving and testing the solutions to the problems of their interests.

\subsection{Parallel computation}
One very important characteristic of \textbf{algorithm} \ref{algo:Time-dependent HJ
    PDE} is that each grid point, within the same time iteration, can be processed
independently to compute a new value in the next time step, and therefore in parallel.
This computational characteristic, in fact, is very desirable for modern multi-threaded
CPU architectures.
To take advantage of this, we need to a way to specify a parallelizable code region in
HeteroCL.
In HeteroCL, this can be achieved by applying the transformation primitive
\textit{parallel} to a loop as illustrated in the snipped code below.

In this code, we iterate through every grid point in a nested for loop fashion.
Within each iteration, we compute the new value at this index according to the target
algorithm.
After specifying the computation procedures, at the end of the code, we apply the
parallel operation to the outmost loop labeled as $i$.
This will signal to subsequent code compilation stage that all index computation are
independent and hence subject to further multi-threading optimization.
Note that for algorithm \ref{algo:Time-dependent HJ PDE}, value function for next time
step is not updated in-place and hence data racing event caused by threads reading and
writing to the same memory block is not an issue.

\begin{lstlisting}[language=Python]
with hcl.Stage("Hamiltonian"):
    with hcl.for_(0, V_init.shape[0], name="i") as i:  
        with hcl.for_(0, V_init.shape[1], name="j") as j:
            with hcl.for_(0, V_init.shape[2], name="k") as k:
                # ...
# Build a computational graph
s = hcl.create_schedule([args], myFunc)
# Choose the computation stage to apply optimization to
s_H = myFunc.Hamiltonian_term
# Parallelize the most outer loop of the stage
s[s_H].parallel(s_H.i)
\end{lstlisting}

At the execution level, available threads is maintained in a thread pool and each
thread member will be assigned a computational tasks from a task queue. In this case,
multiple grid points are assigned to each thread for parallel computation as shown in
\textbf{Figure} \ref{fig:our_implementation}.
The number of threads used equals the maximum number of hyper-threads available in the
CPU.

Note that this parallelization of computation can even be applied to in-place updates
such as value iteration (\textbf{algorithm} \ref{Value Iteration}) and TTR computation
(\textbf{algorithm} \ref{algo: lax_friedrich sweeping}).
For value iteration algorithm and TTR computation, updating multiple grid points
simultaneously might require more iterations until convergence; in certain cases,
however, we find that the overall speedup benefit of parallelization can outweigh the
slight increase of extra iterations.

\subsection{Cache-Aware Loop Iteration  }
\begin{figure}[H]
    \centering
    \includegraphics[width=0.60\textwidth]{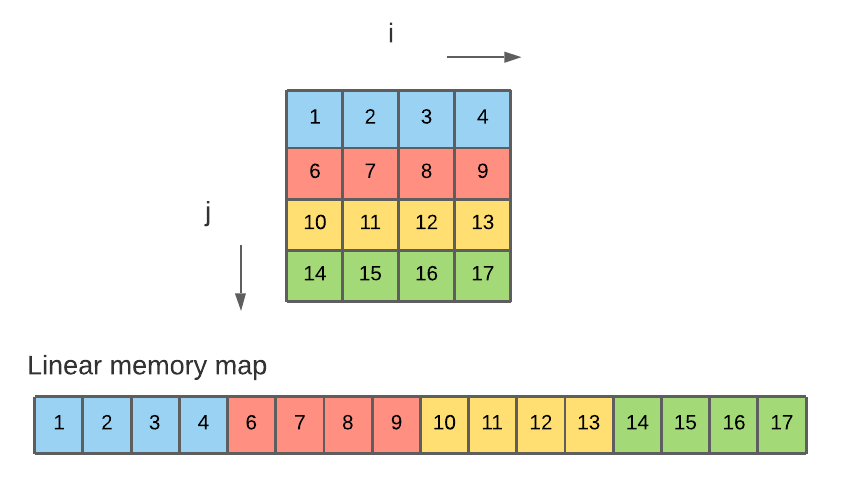}
    \includegraphics[width=0.30\textwidth]{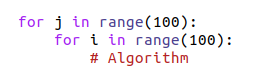}
    \caption{Nested loop order that follows the linear memory map will take advantage of the main memory's spatial locality}

    \label{fig:access_order_cache}
\end{figure}

One important factor that can have a substantial impact on the performance of a program
when dealing with high dimensional arrays is memory locality.
Memory locality refers to the principle that data which are in proximity spatially is
more likely to be retrieved in subsequent operations.
When a memory address $i$ is accessed, data from adjacent address $i+1, i+2, .
    .., i+k$ are loaded onto the local fast memory buffer for fast access by the CPU in the future.
This buffer, known as cache, has very low memory access latency.
If our memory access in the implementation matches well with this caching mechanism, we
can make the most use of this behavior for fast computation.
For example,
in Figure \ref{fig:access_order_cache}, if we iterate the array along the row major order, 
we can reduce the time used loading grid points in subsequent iterations effectively.
This is important for high-dimensional control problem, as the memory access time will
become more dominant when the number of grid points increase.

By knowing the memory layout of the $N$-dimensional array, our grid iterations follow
this layout order which takes advantages of the cache spatial locality.
To abide by Numpy's memory layout, the implementations, by default, assign the highest
dimension being the most inner loop and the lowest dimension being the most outer loop.
Users can define their grid's dimension order so as this nested loop order matches with
the system dynamic's data re-use pattern, which can potentially result in computation
savings.
This optimization applies to all of the algorithm implementations.

\begin{figure}
    \centering
    \includegraphics[width=0.4\textwidth]{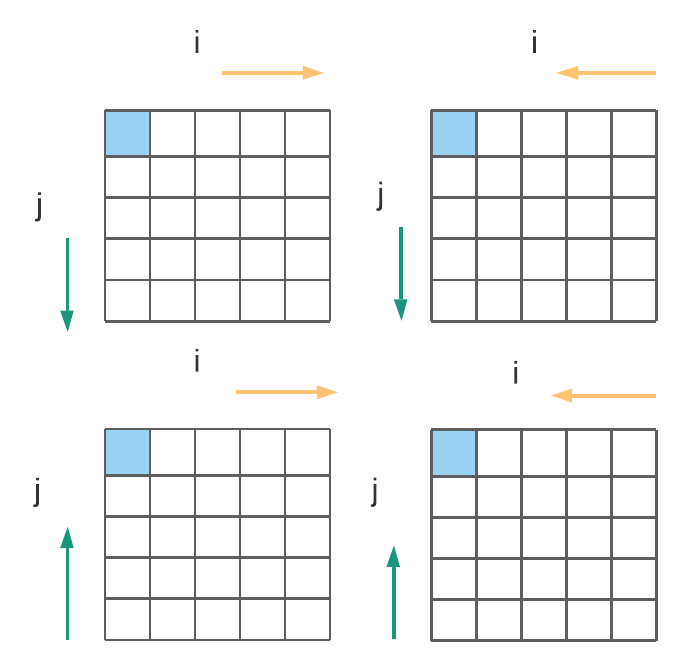}
    \caption{Each grid iteration can have alternating traversing direction for each dimension}
    \label{fig:switch_direction2d}
\end{figure}

\subsection{Alternating sweeping directions }
This optimization is more algorithmic and less on the computer system level, and is
only applicable to in-place updating algorithm such as \textbf{algorithm} \ref{algo:
    lax_friedrich sweeping}, \ref{Value Iteration}.
The general idea of this technique is that certain region of system state space
contains more information than others depending on the dynamics or transition function
of the system.
As such, iterating through these region from a particular direction might result in
faster evaluation of the value function.
Even without knowing beforehand such regions and directions, we can alternate the
directions of iteration in each dimension overtime to exploit this property for faster
value function convergence \cite{Bertsekas1996NeuroDynamic}.
This is illustrated by Figure \ref{fig:switch_direction2d} for a 2-dimensional grid.
In our toolbox, this approach is used in the implementation of value iteration
algorithm and time-to-reach value function.
This technique has been shown to compute time-to-reach value function for 2D systems
\cite{TTR}.

In addition to the optimization that our solver uses, we also attribute the efficiency
of our solver to the compilation workflow of HeteroCL and its underlying backend TVM \cite{chen2018tvm}.
In contrast to Python and MATLAB, implementation of a target algorithm in heteroCL are
compiled ahead of time to generate a computation graph or an intermediate
representation (IR).
This IR is then passed to TVM for further analysis, optimization, scheduling and code
generation.
Such compilation workflow requires memory of multi-dimensional arrays be declared in
advance, which is essential for performant computation of high-dimensional systems.

\section{Benchmarking Results}
\begin{figure}[H]
    \centering
    \includegraphics[width=0.8\textwidth]{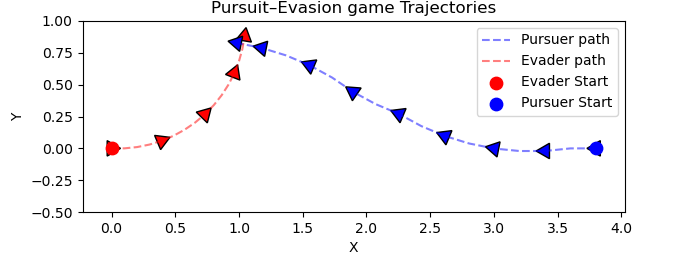}
    \caption{Pursuit-Evasion game trajectory}
    \label{fig:pursuit_evasion_game_plot}
\end{figure}
\begin{table}
    \captionof{table}{Comparisons of computational time (Lower is better) } 
    \begin{center}
        \label{table:dynamic_hj_time_comparisons}
        \begin{tabular}{||c c c c c||}
            \hline \multicolumn{5}{|c|}{\textbf{Computational time (s), time-dependent HJ PDE \cite{HJDynamicGame} ($\downarrow$)} }                                                                                             \\
            \hline
            \textbf{Dimensions}                                   & 3D                              & 4D                                & 5D                              & 6D                             \\
            \hline
            \textbf{Grid points}                                  & $100^3$                         & $60^4$                            & $40^5$                          & $25^6$                         \\
            \hline
            \textbf{Dynamics}                                     & Pursuit Evasion             & Dubins 4D                     & Dubins 5D                   & \shortstack{Underwater \\ Vehicle}             \\
            \hline
            \textbf{Horizon time}                                 & $1.5$                           & $1$                               & $0.3$                           & $20$                           \\
            \hline
            \textbf{Time step}                                    & $0.05$                          & $0.05$                            & $0.05$                          & $0.2$                          \\
            \hline
            \multicolumn{5}{|c|}{\textbf{First Order (Upwind ENO scheme + TVD Runge-Kutta)}}                                                                                                               \\
            \hline
            \textbf{OptimizedDP (Ours)}                                  & $3.2$ (\textbf{$\times$1})      & $29.05$ (\textbf{$\times$1})      & $24.5$ (\textbf{$\times$1})     & $1$ day \\
            \hline
            \textbf{ToolboxLS }\cite{LsetToolbox1}       & $44.8$ (\textbf{$\times$14})    & $455$ (\textbf{$\times$16})       & $806.6$ (\textbf{$\times$33})   & N/A                    \\
            \hline
            \textbf{BEACLS }\cite{BEACLS}                   & $7.0$ (\textbf{$\times$2})      & $57$ (\textbf{$\times $2})        & $87$ (\textbf{$\times$3.6})     & N/A                    \\
            \hline
            \textbf{hj\_reachability }\cite{hjreachability} & $16.49$ (\textbf{$\times $5.2}) & $109.78$ (\textbf{$\times$3.78})  & $368.31$ (\textbf{$\times$15})  & N/A                    \\
            \hline
            \multicolumn{5}{|c|}{\textbf{Second Order (Upwind ENO scheme + TVD Runge-Kutta)} }                                                                                                             \\
            \hline
            \textbf{OptimizedDP (Ours)}                                  & $8.4$ (\textbf{$\times$1})      & $72.6 (\textbf{$\times$1}) $      & $64$ (\textbf{$\times$1})       & $2$ days \\
            \hline
            \textbf{ToolboxLS}\cite{LsetToolbox1}        & $130.36$ (\textbf{$\times $15}) & $1581$ (\textbf{$\times$22 })     & $3152$ (\textbf{$\times$ 49.5}) & N/A                    \\
            \hline
            \textbf{BEACLS }\cite{BEACLS}                    & $18$ (\textbf{$\times$2.14})    & $134$(\textbf{$\times$1.85})      & $213$(\textbf{$\times$3.33})    & N/A                    \\
            \hline
            \textbf{hj\_reachability }\cite{hjreachability} & $67.27$ (\textbf{$\times$16})   & $380.79$ (\textbf{$\times$10.49}) & $1214.93$ (\textbf{$\times$38}) & N/A                    \\
            \hline
        \end{tabular}
    \end{center}
\end{table}

\begin{table}
    \captionof{table}{Comparisons of memory usage (Lower is better) }
    \begin{center}
        \label{table:dynamic_hj_memory_comparisons}
        \begin{tabular}{||c c c c c||}
            \hline \multicolumn{5}{|c|}{\textbf{Memory Usage (GB), time-dependent HJ PDE \cite{HJDynamicGame} ($\downarrow$)}}                                                                                        \\
            \hline
            \textbf{Dimensions}                             & 3D                                & 4D                               & 5D                           & 6D                           \\
            \hline
            \textbf{Grid points}                            & $100^3$                           & $60^4$                           & $40^5$                       & $25^6$                       \\
            \hline
            \textbf{System Dynamics}                        & Pursuit Evasion               & Dubins 4D              & Dubins 5D                   & \shortstack{Underwater \\ Vehicle}             \\
            \hline
            \textbf{Horizon time}                           & $1.5$                             & $1$                              & $0.3$                        & $20$                         \\
            \hline
            \textbf{Time step}                              & $0.05$                            & $0.05$                           & $0.05$                       & $0.2$                        \\
            \hline
            \multicolumn{5}{|c|}{\textbf{First Order (Upwind ENO scheme + TVD Runge-Kutta)}}                                                                                                     \\
            \hline
            \textbf{OptimizedDP (Ours)}                            & 0.1 (\textbf{$\times$1})          & 0.6 (\textbf{$\times$1})         & 4.95 (\textbf{$\times$1})    & 11.33 (\textbf{$\times$1})   \\
            \hline
            \textbf{ToolboxLS}  \cite{LsetToolbox1} & 0.3 (\textbf{$\times$3})          & 3.5 (\textbf{$\times$5.83})      & 28.8 (\textbf{$\times$6.4})  & N/A                  \\
            \hline 
            \textbf{BEACLS} \cite{BEACLS}             & 0.03 (\textbf{$\times$0.3})       & 0.5 (\textbf{$\times$0.83})      & 4.34((\textbf{$\times$0.88}) & N/A                  \\
            \hline
            \textbf{hj\_reachability} \cite{hjreachability} & 0.168 (\textbf{$\times$1.68}) & 2.06 (\textbf{$\times$3.43}) &
            14.8 (\textbf{$\times$2.99})                & N/A                                                                                                                        \\
            \hline
            \multicolumn{5}{|c|}{\textbf{Second Order (Upwind ENO scheme + TVD Runge-Kutta)}}                                                                                                    \\
            \hline
            \textbf{OptimizedDP (Ours)}                     & $0.1$ (\textbf{$\times$1})        & $0.6$ (\textbf{$\times$1})       & $5.45$ (\textbf{$\times$1})  & $12.33$ (\textbf{$\times$1}) \\
            \hline
            \textbf{ToolboxLS} \cite{LsetToolbox1} & 0.3(\textbf{$\times$3})           & 3.6 (\textbf{$\times$6})         & 30.5(\textbf{$\times$5.6})   & N/A                  \\
            \hline
            \textbf{BEACLS} \cite{BEACLS}             & 0.03 (\textbf{$\times$0.3})       & 0.5 (\textbf{$\times$0.83})      & 4.7 (\textbf{$\times$0.86})  & N/A                  \\
            \hline
            \textbf{hj\_reachability} \cite{hjreachability} & 0.143 (\textbf{$\times$1.43})     & 2.2 (\textbf{$\times$3.67})      &
            15.06 (\textbf{$\times$2.76})                   & N/A                                                                                                                        \\
            \hline
        \end{tabular}
    \end{center}
\end{table}

\subsection{Time-Dependent Hamilton-Jacobi PDEs}
In this section, we first compare the performance of optimizedDP against the available
time-dependent HJ PDE implementation in ToolboxLS, hj\_reachability and BEACLS for a various number of
dimensions and problem instances on CPU.
These results are performed on a 16-thread \textbf{Intel(R) Core(TM) i9-9900K CPU} at
3.60GHz with 32 Gigabytes (GBs) of RAM.
Note that, in these results, each column corresponds to different problem instances
with different time-length horizons and minimum stable time steps, which results in
varying computational time as seen in Table \ref{table:dynamic_hj_time_comparisons} and
Table \ref{table:dynamic_hj_memory_comparisons}.
The system dynamics used in the benchmarks can be found in the Supplementary material.

As it can be seen from Table \ref{table:dynamic_hj_time_comparisons} and
\ref{table:dynamic_hj_memory_comparisons}, our toolbox outperforms all other toolboxes on
all problem instances when solving the dynamic HJ PDE in terms of speed.
The only exception is BEACLS \cite{BEACLS} that requires $10-15\%$ less memory than
optimizedDP, which is due to the memory overhead of allocating arrays in
Python. In Fig. \ref{fig:pursuit_evasion_game_plot}, we demonstrate a trajectory of a pursuit-evasion game where the
defender's optimal control is computed from the value function using optimizedDP and it's able to successfully intercept the attacker.

\subsection{Time-Independent Hamilton-Jacobi PDEs}
\begin{figure}
    \centering
    \includegraphics[width=0.6\textwidth]{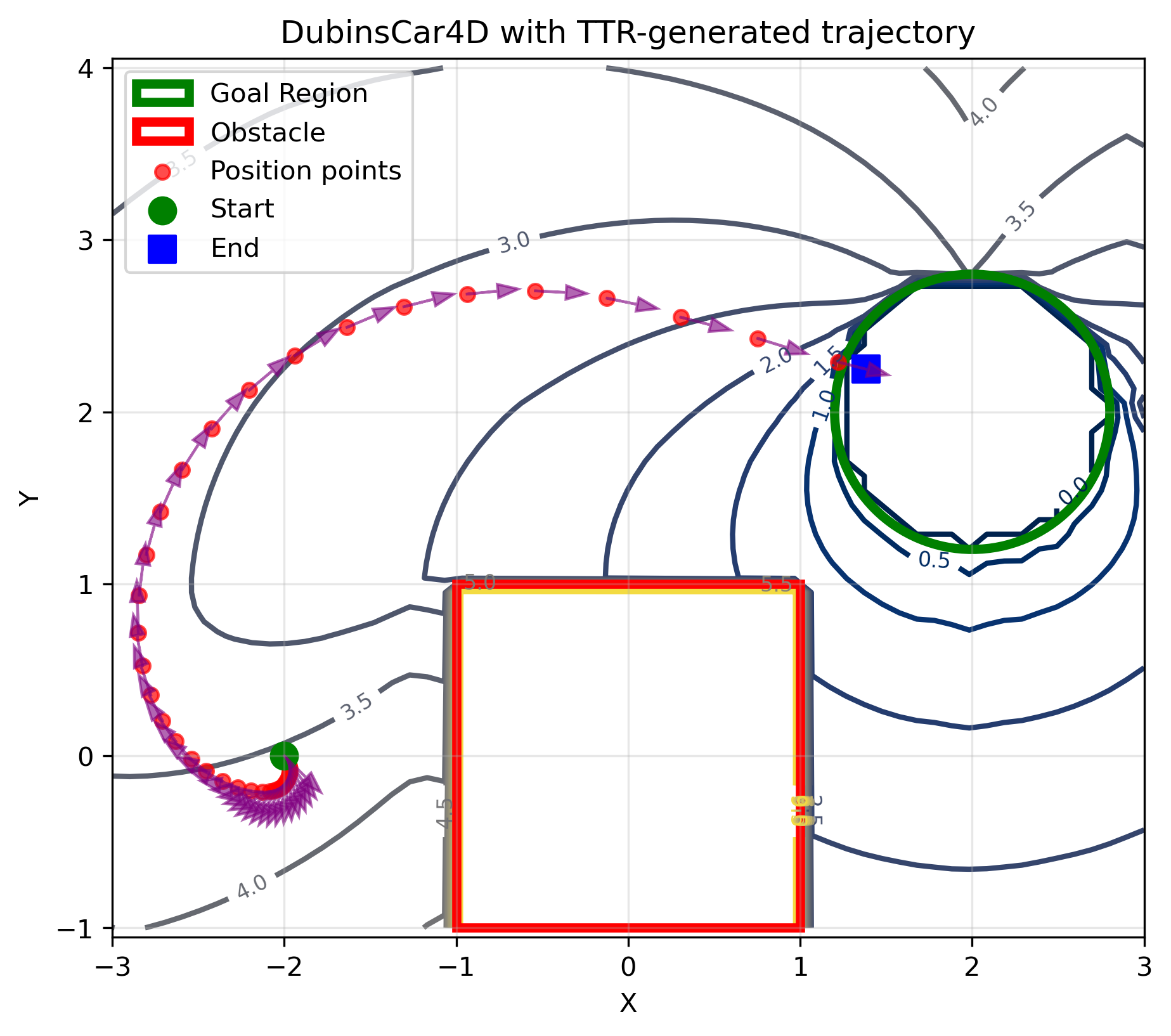}
    \caption{Using computed TTR value by OptimizedDP, a 4D Dubins car can arrive in goal while avoiding obstacles. The contour in
    this figure shows the minimum time to goal. The arrow shows heading of the car.}
    \label{fig:4D-Dubins-car-ttr}
\end{figure}

Since there exists no library that implements algorithm \ref{algo: lax_friedrich sweeping} for the time-independent 
HJ PDE generally for high dimensions, we benchmark our
implementation against naive C++ implementations for 3D systems. 
The comparisons are shown in table \ref{table:Time-to-reach-vs-CPP-results}, which demonstrates faster computational time 
for all grid size. Memory consumption for this test is not reported since it is negligible for both implementations.
We then test our toolbox capabilities on higher dimensions, such as 4D and 6D systems. For these tests,
we gradually increase the number of grid points and record the computational time and memory consumption, which are shown in table \ref{table:Time-to-reach-results4D} and \ref{table:Time-to-reach-results6D}.
We also demonstrate optimal trajectory of a 4D Dubins Car reaching goal and avoiding obstacles
using the computed time-to-reach (TTR) value function in Fig. \ref{fig:4D-Dubins-car-ttr}.


\begin{table}[H]
    \centering
    \begin{minipage}[t]{0.32\textwidth}
        \captionof{table}{3D Dubins Car}
        \label{table:Time-to-reach-vs-CPP-results}
        \begin{tabular}{||c c c c||}
            \hline \multicolumn{4}{|c|}{\textbf{Time (s)}} \\ \hline
            \textbf{Grid} & $60^3$ & $80^3$ & $100^3$ \\ \hline
            Ours & $0.31$ & $0.65$ & $1.41$ \\ \hline
            C++ & $1.02$ & $3.02$ & $6.85$ \\ \hline
        \end{tabular}
    \end{minipage}
    \hfill
    \begin{minipage}[t]{0.32\textwidth}
        \captionof{table}{4D Dubins Car}
        \label{table:Time-to-reach-results4D}
        \begin{tabular}{||c c c c||}
            \hline \multicolumn{4}{|c|}{\textbf{Time (s)}} \\ \hline
            \textbf{Grid} & $60^4$ & $80^4$ & $100^4$ \\ \hline
            Ours & $12$ & $43.3$ & $125.8$ \\ \hline
            \multicolumn{4}{|c|}{\textbf{Memory (GB)}} \\ \hline
            Ours & 0.3 & 0.94 & 2.29 \\ \hline
        \end{tabular}
    \end{minipage}
    \hfill
    \begin{minipage}[t]{0.32\textwidth}
        \captionof{table}{6D Dual Dubins Car}
        \label{table:Time-to-reach-results6D}
        \begin{tabular}{||c c c c||}
            \hline \multicolumn{4}{|c|}{\textbf{Time (s)}} \\ \hline
            \textbf{Grid} & $10^6$ & $20^6$ & \shortstack{$30^4$ \\ $\times 20^2$} \\ \hline
            Ours & $8.9$ & $207$ & $1501$ \\ \hline
            \multicolumn{4}{|c|}{\textbf{Memory (GB)}} \\ \hline
            Ours & 0.049 & 2.5 & 13.2 \\ \hline
        \end{tabular}
    \end{minipage}
\end{table}

\subsection{Value Iteration}
\begin{figure}
    \centering
    \begin{subfigure}{.25\textwidth}
        \centering
        \includegraphics[width=\linewidth]{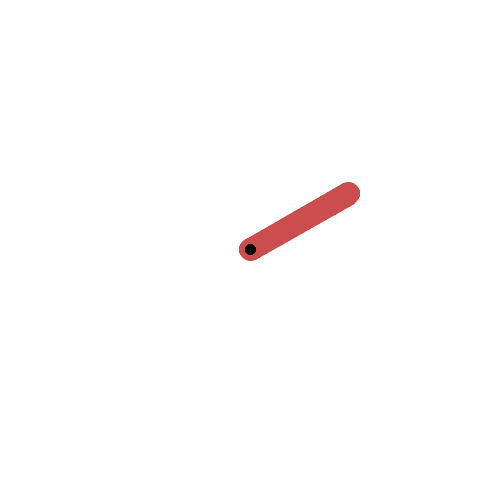}
        \caption{Step 0}
    \end{subfigure}\hfill
    \begin{subfigure}{.25\textwidth}
        \centering
        \includegraphics[width=\linewidth]{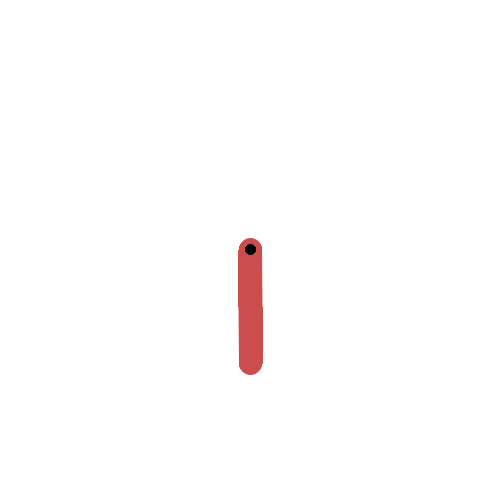}
        \caption{Step 10}
    \end{subfigure}\hfill
    \begin{subfigure}{.25\textwidth}
        \centering
        \includegraphics[width=\linewidth]{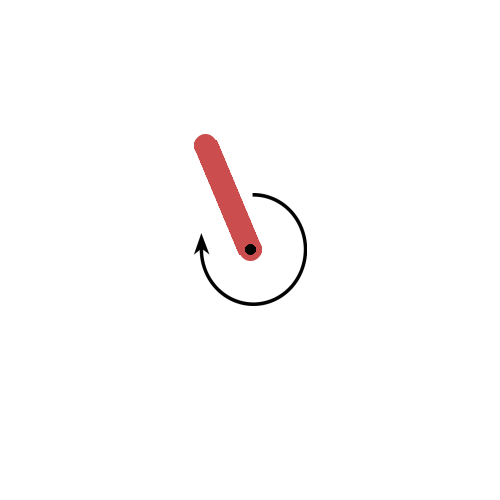}
        \caption{Step 30}
    \end{subfigure}\hfill
    \begin{subfigure}{.25\textwidth}
        \centering
        \includegraphics[width=\linewidth]{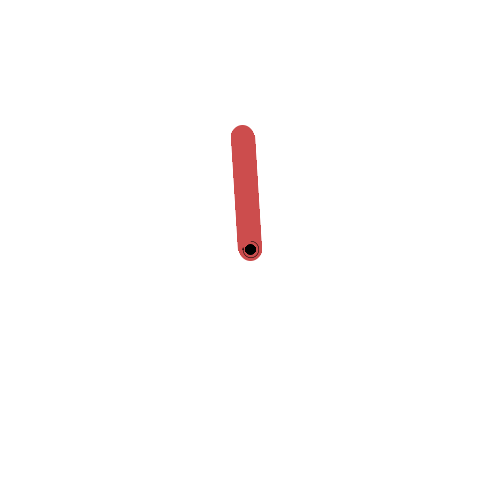}
        \caption{Step 60}
    \end{subfigure}
    \caption{ An inverted pendulum is successfully balanced using the optimal policy computed from value iteration using OptimizedDP}
    \label{fig:inverted_pendulum_working}
\end{figure}

Similarly, since there exists no standard toolbox for solving discretized continuous value iteration, 
we compare OptimizedDP against implementation in Python.
In this section, we apply discretized value iteration to different openAI gym environments
with continuous domains.
With optimizedDP, we are able to compute optimal value function to balance a 2D inverted
pendulum (Fig. \ref{fig:inverted_pendulum_working}) in less than 10 seconds without a GPU, which is shown in Table \ref{table:value-iteration-results-2D}
and much faster than most common model-free reinforcement learning algorithms.

We also show that OptimizedDP can perform value iteration large problem instances that would be intractable for Python such as 
the 4D Cartpole and 6D planar quadrotor. The computational time of solving these problems are shown in 
Table \ref{table:value-iteration-results-4D} and \ref{table:value-iteration-results-6D}. For these problems, although the memory usage of the
Python program is tractable, the computation doesn't seem to make progress or finish in a reasonable amount of time. Additionally,
we notice that as the grid size becomes bigger, the computational time start to increase significantly at some point
even though the memory usage increase linearly as expected, which is observed in the 6D planar quadrotor problem.
This is because a finer grid would require more iterations to converge to the optimal value function with a 
smaller time step $\Delta t$, while each iteration
takes longer time to compute. Even though value iteration requires knowing the transitions matrix and the reward
function, this result proves that there are sufficient
computation power for tractable updates of the Bellman equation to obtain optimal control for high-dimensional
control problems.

\subsection{High-dimensional stress-testing}

\begin{figure}[ht]
    \centering
    \includegraphics[width=0.80\textwidth]{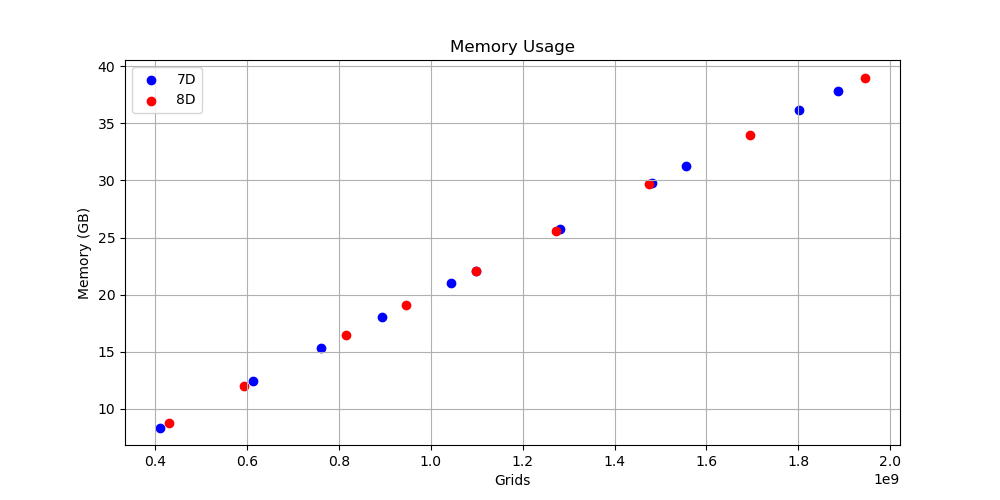}
    \caption{RAM consumptions increase linearly as a function of number of grid points}
    \label{fig:Memory_7D8D}
\end{figure}

\begin{figure}[ht]
    \centering
    \includegraphics[width=0.80\textwidth]{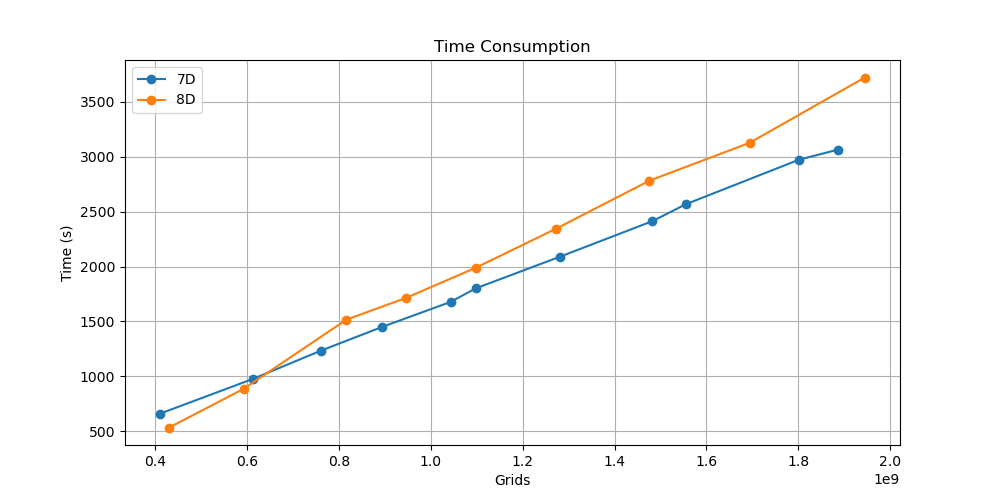}
    \caption{Computational time of solving time-dependent HJ pde for 7 and 8 dimensions increases linearly as a function of grid points}
    \label{fig:Time_7D8D}
\end{figure}

In this section, we show that, given larger RAM capability such as a compute server, our toolbox can be utilized to solve
reachability problems for even bigger high-dimensional problems shown in previous section. 
To stress-test and demonstrate this capability, we solve time-dependent HJ PDEs on a multi-core server
machine equipped with 1TB of RAM for dynamical systems of 7 and 8 dimensions. Since solving the 
time-dependent HJ PDE is a more memory and computationally intensive process than solving the time-independent HJ PDE and
value iteration given the same grid size, the results in this section can also be extrapolated to
these two algorithms implementation. As we vary the number of grid points, 
we record that the memory usage and computational time of solving the time-dependent HJ PDE.
The results of memory usage and computational time of these experiments are
shown in Fig.
\ref{fig:Memory_7D8D} and Fig.\ref{fig:Time_7D8D} respectively.
In these experiments, the artificial dissipation coefficients $\alpha_i$ in each dimension of
Algorithm \ref{algo:Time-dependent HJ PDE} are approximated as the maximum rate of changes 
over the Cartesian grid to help reduce computation time and memory usage. It can observed from Table
\ref{fig:Memory_7D8D} that RAM usage is independent of the number of dimensions and only depends on the number of grid points.
And from Table \ref{fig:Time_7D8D}, it can be observed that $8$D problem incur slightly more computational time than
$7$D problem given the same number of grid points, which is because of to a smaller CFL time step size resulting
in more iterations to integrate.

\begin{table}[H]
    \captionof{table}{2D inverted pendulum (Value Iteration)}
    \begin{center}
        \label{table:value-iteration-results-2D}
        \begin{tabular}{||c c c c||}
            \hline \multicolumn{4}{|c|}{\textbf{Computational time (seconds)} }                                                            \\
            \hline
            \textbf{$|S| \cross |A|$} & $37\times81\times21$   & $73\times81\times21$   & $73\times163\times42$ \\
            \hline
            \textbf{OptimizedDP (Ours)}      & $1.9$           & $3.11$          & $9.37$         \\
            \hline
            \textbf{Python}           & $2360$          & $3227$          & $7602$         \\
            
            \hline \multicolumn{4}{|c|}{\textbf{Memory Usage (Gigabytes)} }\\
            \hline
            \textbf{OptimizedDP (Ours)}      & $0.15$           & $0.15$          & $0.15$         \\
            \hline
            \textbf{Python}           & $0.04$          & $0.04$          & $0.04$         \\
            \hline                                                            
        \end{tabular}
    \end{center}
\end{table}

\begin{table}[H]
    \centering
    \begin{minipage}[t]{0.48\textwidth}
        \captionof{table}{4D Cartpole (Value Iteration)}
        \label{table:value-iteration-results-4D}
        \begin{tabular}{||c c c c||}
            \hline \multicolumn{4}{|c|}{\textbf{Time (s)} } \\
            \hline
            \shortstack{$|S|$ \\ $\cross$ \\ $|A|$} & \shortstack{$200\times50$ \\ $\times40\times50$ \\ $\times2$}   & \shortstack{$200\times100$ \\ $\times40\times50$ \\ $\times2$}   & \shortstack{$200\times100$ \\ $\times80\times50$ \\ $\times2$} \\
            \hline
            \shortstack{Ours}      & $47.25$           & $93$          & $182$         \\
            \hline \multicolumn{4}{|c|}{\textbf{Memory (GB)} }\\
            \hline
            \shortstack{Ours}      & $0.46$           & $0.77$          & $1.37$         \\
            \hline
        \end{tabular}
    \end{minipage}%
    \hfill
    \begin{minipage}[t]{0.48\textwidth}
        \captionof{table}{6D planar quadrotor (Value Iteration)}
        \label{table:value-iteration-results-6D}
        \begin{tabular}{||c c c c||}
            \hline \multicolumn{4}{|c|}{\textbf{Time (s)} } \\
            \hline
            \shortstack{$|S|$ \\ $\cross$ \\ $|A|$} & \shortstack{$40^2\times15^2$ \\ $\times36\times10$ \\ $\times160$} & \shortstack{$40^2\times20^2$ \\ $\times36\times10$ \\ $\times160$} & \shortstack{$50^2\times20^2$ \\ $\times36\times15$ \\ $\times160$}  \\
            \hline
            \shortstack{Ours}      & $4400$           & $5383$          & $65440$         \\
            \hline \multicolumn{4}{|c|}{\textbf{Memory (GB)} }\\
            \hline
            \shortstack{Ours}      & $4.5$           & $8.1$          & $16$         \\
            \hline
        \end{tabular}
    \end{minipage}
\end{table}

\section{Limitation and future work}
We have shown that, given enough computational resources, backup-based optimization can
be performed for high-dimensional control problem that are considered intractable
before.
Although the toolbox will not solve the ``curse of dimensionality", we believe the
toolbox, in effective combination with dimension reduction methods
\cite{DecomposeMethod}, warm-up techniques \cite{InitializationWarmUp}, and learning
methods, can solve more interesting control problems.
Finally, OptimizedDP toolbox is still a work in progress and we plan on adding new features
to the toolbox such as higher order ENO scheme for more accurate derivatives approximation.


\begin{acks}
To Robert, for the bagels and explaining CMYK and color spaces.
\end{acks}

\bibliographystyle{ACM-Reference-Format}
\bibliography{references}

\appendix
\section{Appendices}
\label{appendix}

This Appendix contains details about each of the examples we ran for our benchmarks.

\textbf{2D Double Integrator} system dynamics:

\begin{equation}
    \label{eq:quadrotor}
    \begin{aligned}
        \dot{x} & = v_x & \dot{v}_{x} & = u_x \\
        \dot{y} & = v_y & \dot{v}_{y} & = u_y
    \end{aligned}
\end{equation}
where $ \abs{x}, \abs{y} \leq 2.0 $ are the positions, $ \abs{v_x}, \abs{v_y} \leq 2.0$ are the velocity in the the two dimensions, and $\abs{u_x}, \abs{u_y} \leq 1$ are the two inputs.

\textbf{3D Pursuit and Evasion} system dynamics:
\begin{equation}
    \label{eq:car_dynamics}
    \begin{aligned}
        \dot{x}      & = -v_{a} + v_{b}\cos\theta + ay \\
        \dot{y}      & = v_a \sin\theta - ax           \\
        \dot{\theta} & = b-a
    \end{aligned}
\end{equation}
where $\abs{x} \leq 4$, $\abs{y} \leq 4$, $ -\pi \leq \theta < \pi$ are the relative positions and heading, $v_a = 1$ and $v_b = 1$ are the evaders and pursuer's speed, $\abs{a} \leq 1$ and $\abs{b} \leq 1$ are the control input of the evader and pursuer respectively.

\textbf{3D Dubins Car} system dynamics:
\begin{equation}
    \label{eq:car_dynamics}
    \begin{aligned}
        \dot{x}      & = v\cos\theta \\
        \dot{y}      & = v\sin\theta \\
        \dot{\theta} & = \omega
    \end{aligned}
\end{equation}
where $-3.0 \leq x \leq 3.0 $, $-1.0 \leq y \leq 4.0 $, $ -\pi \leq \theta < \pi$ are the positions and heading respectively, $v = 1$ is the constant speed and $\abs{\omega} \leq 1.0 $ is the input angular acceleration.

\textbf{4D Extended Dubins Car} system dynamics:

\begin{equation}
    \label{eq:extendedDubins4d}
    \begin{aligned}
        \dot{x} & = v  \cos(\theta) & \dot{y}      & = v \sin(\theta)          \\
        \dot{v} & = a               & \dot{\theta} & = v\frac{\tan(\delta)}{L}
    \end{aligned}
\end{equation}
where $-3 \leq x \leq 3$, $-1 \leq y \leq 4$ are the positions, $ 0 \leq v \leq 4 $ is the speed, $ -\pi \leq \theta < \pi$ is the orientation,  $-1.5 \leq a \leq 1.5 $ and $-\pi/15 \leq  \delta  < \pi/15$ are the control inputs.

\textbf{4D Dubins Car} system dynamics:

\begin{equation}
    \label{eq:Dubins4D}
    \begin{aligned}
        \dot{x} & = v  \cos(\theta) & \dot{y}      & = v \sin(\theta) \\
        \dot{v} & = a               & \dot{\theta} & = \omega
    \end{aligned}
\end{equation}
where {$a, \omega$} are the control inputs.

\textbf{5D Dubins Car} system dynamics:
\begin{equation}
    \label{eq:car_dynamics2}
    \begin{aligned}
        \dot{x}      & = v  \cos(\theta) & \dot{y}      & = v \sin(\theta) \\
        \dot{v}      & = a               & \dot{\theta} & = \omega         \\
        \dot{\omega} & = u
    \end{aligned}
\end{equation}
where $a, u$ are the control inputs.

\textbf{6D Underwater vehicle} system dynamics \cite{UnderwaterPaper}:
\begin{align}\label{eqn:state-dependent-relative-dynamics}
    \dot{x}_{\alpha} & = u_r + {V}_{f,x}(x,z,t) + d_x - b_x \nonumber                      \\
    \dot{z}_{\alpha} & = w_r + {V}_{f,z}(x,z,t) + d_z - b_z \nonumber                      \\
    \dot{u}_r        & = \frac{1}{m-X_{\dot{u}}}((\bar{m} - m)A_{f,x}(x,z,t) \nonumber     \\
                     & \quad - (X_u + X_{\abs{u} u}\abs{u_r})u_r +  T_A) + d_u \nonumber   \\
    \dot{w}_r        & = \frac{1}{m-Z_{\dot{w}}} ((\bar{m} - m)A_{f,z}(x,z,t) \nonumber    \\
                     & \quad -(-g(m-\bar{m})) - (Z_w + Z_{\abs{w}w}\abs{w_r})w_r \nonumber \\
                     & \quad  + T_B) + d_w                                                 \\
    \dot{x}          & = u_r + {V}_{f,x}(x,z,t) + d_x\nonumber                             \\
    \dot{z}          & = w_r + {V}_{f,z}(x,z,t) + d_z\nonumber
\end{align}
where $x, z$ denote the vehicle position, $u_r, w_r$ represent relative velocities between vehicle and water flow, $x_{\alpha}, z_{\alpha}$ denote relative position between tracker and planner.
The control inputs are $T_A, T_B$, planning inputs are $b_x, b_z$, and disturbances are
$d_x, d_z, d_u, d_w$.
The problem parameters are $m, \bar{m}, X_{\dot{u}}, Z_{\dot{w}}, X_u, X_w$, $X_{|u|u}, Z_{|w|w}$.

\textbf{6D Planar Quadrotor} system dynamics:
\begin{align}\label{eqn:planar_quadrotor}
    \dot{x} & = v_x \nonumber\\
    \dot{z} & = v_z \nonumber\\
    \dot{v}_x & = -u_T \sin \theta \nonumber \\
    \dot{v}_z & = u_T \cos \theta - g \nonumber\\
    \dot{\theta} & = \omega \nonumber \\
    \dot{\omega} & = u_\tau
\end{align}

\noindent where $x, z$ are the positions, $v_x, v_z$ are the velocities, 
$\theta$ is pitch angle, $\omega$ is pitch rate, $u_T$ is the thrust input, 
and $u_\tau$ is the torque input.

\end{document}